\documentclass[prd,letterpaper,twocolumn,tightenlines,superscriptaddress,showpacs,preprintnumbers,nofootinbib,floatfix,psfig,showpacs,showkeys]{revtex4-1}

\usepackage{amsmath,amssymb,amsfonts,graphicx,pifont,dcolumn}
\usepackage{epstopdf}
\usepackage{color}
\usepackage{cancel}

\newcommand{\bit}{\begin{itemize}}
\newcommand{\eit}{\end{itemize}}

\def\be{\begin{equation}}
\def\ee{\end{equation}}
\def\bea{\begin{eqnarray}}
\def\eea{\end{eqnarray}}

\begin{document}




\twocolumngrid
\title{Generalized Parton Distributions: \\ confining potential effects within AdS/QCD}

\author{Marco~Traini} 
\address{Institut de Physique Th\'eorique, Universit\'e Paris Saclay, CEA, F-91191 Gif-sur-Yvette, France}
\address{INFN - TIFPA, Dipartimento di Fisica, Universit\`a degli Studi di Trento,\\ 
Via Sommarive 14, I-38123 Povo (Trento), Italy}

\begin{abstract}
Generalized Parton Distributions are investigated within a holographic approach where the string modes in the fifth dimension describe the nucleon in a bottom-up or AdS/QCD framework. The aim is to bring the AdS/QCD results in the realm of phenomenology in order to extract consequences and previsions. Two main aspects are studied: i) the role of the confining potential needed for breaking conformal invariance and introducing confinement (both: classic Soft-Wall and recent Infra-Red potentials are investigated); ii) the extension of the predicted GPDs to the entire range of off-forward kinematics by means of Double-Distributions. Higher Fock states are included describing the nucleon as a superposition of three valence quarks and quark-antiquark pairs and gluons.

\end{abstract}

\maketitle

\section{\label{sec:intro}Introduction}

Generalized Parton Distribution functions (GPDs) are a source of fundamental information encoding  essential aspects of the nucleon structure \cite{DiehlPR2003,generalGPDs,GPDs_perp,GPDs_FF,GPDs_structure,GPDs_q}
as basic ingredients in the description of hard exclusive processes \cite{generalGPDs}. They are generalization of the well known parton distribution functions and, at the same time, as correlation functions they incorporate quite non trivial aspects of hadrons in the non-perturbative regime like: electromagnetic form factors, spin and angular momentum of the constituents and their spatial distribution \cite{DiehlPR2003,GPDs_FF}.
Their functional structure is usually written as function of the longitudinal momentum fraction of the active quark ($x$), the momentum transferred  in the longitudinal direction ($\xi$ or skewedness) and the invariant momentum (square) $t = - \Delta^2$.
The Fourier transform of GPDs (at $\xi=0$) in the transverse direction encodes information on the partonic  distributions in the transverse plane and it translates in a quantitative information (because of the probabilistic interpretation as density functions) on the separation of the struck quark and the center of momentum of the nucleon \cite{GPDs_perp}. The detailed map of quarks and gluons in the nucleon interior is often called "nucleon tomography" since the traditional information from elastic and deep-inelastic scattering provide static coordinates or momentum space pictures, separately, while  GPDs provide pictures of dynamical correlations in both coordinate and momentum spaces \cite{XJi_prize}.

Amplitudes of different hard exclusive processes (like deeply virtual Compton scattering (e.g. \cite{GPDs_CLAS_VCSc2015}), and virtual vector meson production (e.g. \cite{GPDs_CLAS_VMProd}) in the new generation of CLAS experiments at Jefferson Lab.) contain GPDs as essential components. On the other hand the experimental results already collected have shed a fundamental light on their role in different processes and kinematical regimes (for example the H1 \cite{GPDs_H1} and ZEUS \cite{GPDs_ZEUS} at HERA, HERMES at DESY \cite{GPDs_HERMES}, Hall A and Hall B at Jefferson Lab. \cite{GPDs_HallA}, COMPASS at CERN  \cite{GPDs_COMPASS}).

GPDs are non-perturbative objects and their evaluation lies in the realm of non-perturbative QCD. The successes are, till now, strongly limited \cite{GPDs_Lat}. An alternative approach is the Holographic Light Front technique. Its fundamentals are in the correspondence between string theory developed in a higher dimensional ant-de-Sitter (AdS) space and conformal field theory (CFT) in Minkowski physical space-time \cite{AdS_Maldacena,AdS_Polyakov,AdS_Witten1and2}. Several consequent models have been constructed and they can be divided in top-down and bottom-up approaches. Starting from some brane configuration in string theory, one can, indeed, try to reproduce basic features of QCD following top-down paths (e.g. ref. \cite{top_down}). On the way up one starts from low-energy properties of QCD (like chiral symmetry breaking and quark confinement) to infer elements for a gravity frame with asymptotically anti-de Sitter (AdS) space, the models are therefore indicated as AdS/QCD (e.g. ref.\cite{hep-ph/1407.8131,BT1} and references therein). In particular within the bottom-up approach two successful models have been constructed: i) the Hard-Wall model which uses a sharp cut-off in the extra dimension to confine the (dual) hadron field \cite{HW1,HW2}. The model is simple, analytic and appealing, but it does not reproduce the linear Regge behavior of the meson masses; ii) in the Soft-Wall model \cite{SW1} a (quadratic) dilation field is added in the meson sector in order to successful reproduce the Regge behavior, however chiral symmetry breaking cannot be consistently realized. In particular it has been shown \cite{ChiralTransition_SW2016} that the spontaneous chiral symmetry breaking in vacuum and its restoration at finite temperature, can be realized only within a careful choice of the dilaton profile (see also ref.\cite{RealizationChiSB2016}).

Consequently several authors are investigating how to improve the SW description to incorporate the largest number of QCD properties \cite{IR_isospectral,IR_improvedN,IR_improvedM,IR_glueball,IR_dynamics}.

An example particularly interesting in the present perspective is the Infra-Red improved soft-wall AdS/QCD model proposed in ref. \cite{IR_improvedN}: it is constructed for baryons,  taking into account a specific baryonic property of the spectrum, namely the parity-doublet pattern of the excited baryons. It shows consistent properties also in the meson sector \cite{IR_improvedM}. This simplified model is taken, in the present paper,  as a prototype to investigate GPDs and illustrating, at the same time, a procedure valid to study Generalized Parton Distributions and other observables  in a generically modified confining potential.

Within the AdS/QCD  approach Deep Inelastic Scattering (DIS) has been first addressed by Polchinski and Strassler in refs. \cite{AdS_DIS}, and GPDs have been investigated by many authors
both within the Hard-Wall \cite{GPDs_HW} and Soft-Wall \cite{GPDs_SW,GPDs_SW2} models. Because of the nature of the AdS-QCD analogy in the region of DIS, the results are restricted to the {\it forward} limit ($\xi = 0$) (cfr. Section \ref{sec:GPDs-SR}).

In the present work an attempt for a step forward is investigated and in two directions: i) generalizing the study of GPDs for confining potentials more complex than the simple Soft-Wall model; ii) extending the GPDs results to the off-forward region, $\xi > 0$, by means of a technique called Double-Distributions 
\cite{two_component1999}.

In Section \ref{sec:SWtoIR} the procedure to evaluate the nucleon holographic wave function in the modified confining potential is discussed and the numerical results illustrated. In Section \ref{sec:GPDs-SR}
the relation between sum rules and the $\xi=0$ components of the GPDs is investigated and generalized to include, within a unified framework: i) the effects of the modified confining potential; ii) the contributions of higher Fock states. Numerical results for both helicity-independent and -dependent GPDs  are discussed in Section \ref{sec:results} and compared with a Light-Front approach. Section \ref{sec:modeling} is devoted to the application of Double-Distribution techniques \cite{two_component1999} to the AdS/QCD predictions for the Soft-Wall model. It is shown how AdS/QCD can become predictive in the whole kinematical range $(x,\xi>0,t)$.
Conclusions and perspectives  in Section \ref{sec:conclusions}.

\section{\label{sec:SWtoIR}From the Soft-Wall to the Infra-Red improved model}

The AdS/QCD framework relates a gravitationally interacting theory in the ant-de-Sitter space AdS$_{d+1}$ with a conformal gauge theory in $d$-dimensions defined at the boundary. The needed breaking of conformal invariance (QCD is not a conformally invariant theory) of that correspondence for the baryonic case is obtained introducing, in addition to the dilaton term $\varphi(z)$, an effective interaction $\rho(z)$ in the action of the Dirac field (propagating in AdS$_{d+1}$) \cite{hep-ph/1407.8131,BT1}:
\bea
S & =& {1 \over 2} \int d^dx \, dz \, \sqrt{g}\, e^{\varphi(z)} \times \nonumber \\
&\times& \left[\Psi \left(i\Gamma^A e^M_A D_M - \mu - \rho (z)\right) \Psi  + h.c. \right]\,.
\label{eq:F-action}
\eea
Maximal symmetry is restored for $\varphi(z) = \rho(z) = 0$. One has $\sqrt(g) = \left({R \over z}\right)^{d+1}$ while $e^M_A$ is the inverse vielbein, $e^M_A = \left({z \over R}\right) \delta^M_A$.  $D_M$ is the covariant derivative and the Dirac matrices anti-commute  $[\Gamma^A,\Gamma^B] = 2 \eta^{AB}$. A Dirac-like wave equation can be derived from Eq.(\ref{eq:F-action}) and the dynamical effect due to the dilaton field reabsorbed rescaling the spinor
$\Psi \to e^{\varphi(z)/2} \Psi$. For that reason the term $e^{\varphi(z)}$ is sufficient to break maximal symmetry for mesons but not for the baryon sector. The additional interaction term $\rho(z)$ provides the needed breaking (and confining) contributions to generate the correct baryon spectrum \cite{hep-ph/11080346,hep-ph/1407.8131}. The absence of dynamical effects of the dilaton background field has a particular disappointing  side effect in the lack of guidance from gravity to solve the equations.

A solution is given by a Light-Front holographic mapping where the LF wave equation can be identified with the equation of motion. In the case of $d=4$, $\Gamma^A = (\gamma_\mu, i\gamma_5)$ and $V(z) = \left({R \over z}\right) \rho(z)$, the holographic variable $z $ can be identified with the transverse impact variable $\zeta$ of the $n-1$ spectator system with respect the active parton in a $n$-parton bound state ($z = \zeta$). In $2 \times 2$ chiral spinor representation one obtains two coupled differential equations (e.g. ref. \cite{hep-ph/1407.8131})
\bea
{d \over d\zeta} \phi^+ - {\nu+1/2 \over \zeta} \phi^+ - V(\zeta) \phi^+ & = &{\cal M} \phi^-\,,
\label{eq:psiplusminus}\\
-{d \over d\zeta} \phi^- - {\nu+1/2 \over \zeta} \phi^- - V(\zeta) \phi^-  & = & {\cal M} \phi^+\,;
\label{eq:psiminusplus}
\eea
where $\nu$ can be identified with the light-front angular momentum, i.e. the relative angular momentum between the active parton and the spectator cluster.
Eqs.(\ref{eq:psiplusminus}) and (\ref{eq:psiminusplus}) are easily reduced to the equivalent system of second order differential equations: 
\bea
&-&{d^2 \over d\zeta^2} \phi^+ - {1-4\nu^2 \over 4 \zeta^2} \phi^+ + {2\nu+1 \over \zeta} V(\zeta) \phi^+ + \nonumber \\
&+& {d V(\zeta)\over d \zeta} \phi^+ + V^2(\zeta) \phi^+  = {\cal M}^2 \phi^+\,,
\label{eq:psiplus}
\eea
\bea
&-&{d^2 \over d\zeta^2} \phi^- - {1-4(\nu+1)^2 \over 4 \zeta^2} \phi^- + {2\nu+1 \over \zeta} V(\zeta) \phi^- +
\nonumber \\
&-&{d V(\zeta)\over d \zeta} \phi^- + V^2(\zeta) \phi^-  = {\cal M}^2 \phi^- \,. \label{eq:psiminus}
\eea

\subsection{\label{sec:SW-linear} Linear Soft-Wall potential}

For a quadratic interaction (and $z=\zeta$ within the holographic model), $\rho(\zeta) \sim \zeta^2$, $V(\zeta) = \alpha^2 \zeta$ (the so called Soft-Wall linear potential) and Eqs.(\ref{eq:psiplus}) (\ref{eq:psiminus}) become:
\bea
&-&{d^2 \over d\zeta^2} \phi^+ - {1-4\nu^2 \over 4 \zeta^2} \phi^+  + \alpha^4 \zeta^2  \phi^+ + 2(\nu+1) \alpha^2 \phi^+ = \nonumber \\
&=& {\cal M}^2 \phi^+\,, \label{eq:psipluslinear} \\
&-&{d^2 \over d\zeta^2} \phi^- - {1-4(\nu+1)^2 \over 4 \zeta^2} \phi^-  + \alpha^4 \zeta^2 \phi^-  + 2 \nu \alpha^2 \phi^- = \nonumber \\
&=& {\cal M}^2 \phi^- \,. 
\label{eq:psiminuslinear}
\eea
with normalized solutions (equivalent to $2D$ - harmonic oscillator)
\bea
\phi^+_{n,l_+}(\zeta) & = & \sqrt{2 \, n! \over (n+l_+) !}\sqrt{\alpha} \,(\alpha \zeta)^{l_++1/2} \, e^{-\alpha^2 \zeta^2/2} \, L^{l_+}_n(\alpha^2 \zeta^2), \nonumber \\ \label{eq:sol-psipluslinear}\\
\phi^-_{n,l_-} (\zeta) & = & \sqrt{2 \, n! \over (n+l_-) !}\sqrt{\alpha}\,(\alpha \zeta)^{l_-+1/2} \, e^{-\alpha^2 \zeta^2/2} \, L^{l_-}_n(\alpha^2 \zeta^2); \nonumber \\ \label{eq:sol-psiminuslinear}
\\ \nonumber
\eea
where
\bea
\int d\zeta\,|\phi^+_{n, l_+}(\zeta)|^2 = \int d\zeta\,|\phi^-_{n,l_-}(\zeta)|^2 = 1\,.
\eea
$L_n^l(x)$ are the associated Laguerre polynomials and  one identifies common
eigenvalues ${\cal M}^2 = 4 \alpha^2 (n+\nu+1)$.  The linear confining potential generates a mass gap of the order of $\alpha$. $\nu$ is related to the h.o. angular momentum by $l_+ = \nu$,  $l_- = \nu+1$. In the following $\alpha^2 =(0.41)^2$ GeV$^2$ will be selected, a value which interpolates among different choices in the literature (cfr. ref.\cite{hep-ph/1407.8131}  and references therein) and it gives a good fit to the Form factors \cite{GPDs_SW2}. A critical analysis of the influence of the $\alpha$'s value on the results of the present approach will be given in Section \ref{sec:alpha}. 

\subsection{\label{sec:SW-IR}The IR-improved Soft-Wall model and its solutions}

\begin{figure}[tbp]
\centering\includegraphics[width=\columnwidth,clip=true,angle=0]{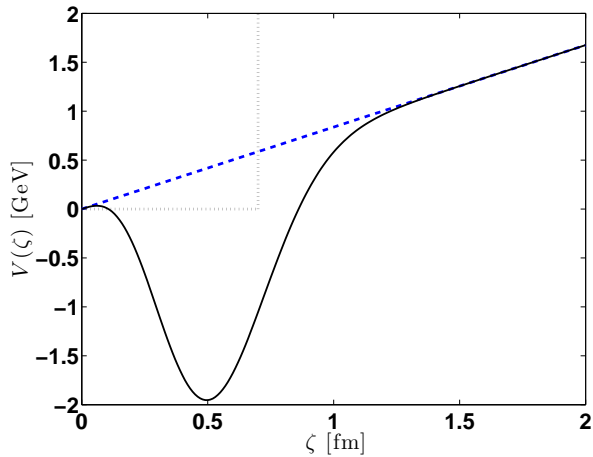}
\caption{\small (Color online) The confining Soft-Wall linear potential $V(\zeta) = \alpha^2 \zeta$ ($\alpha = 0.41$ GeV) as function of $\zeta$ (fm) (dashed line), is compared with the IR-improved potential introduced in refs.\cite{IR_improvedM,IR_improvedN} (see Eq.(\ref{eq:Vconf_IR}) and Section \ref{sec:SW-IR} for comments). Also a hard-wall potential at $\zeta_0 \sim 1/\Lambda_{\rm QCD}$ is sketched (dotted line).}
\label{fig:Vconf_IR}
\end{figure}

The Infra-Red improved Soft-Wall AdS/QCD model proposed in ref. \cite{IR_improvedN} (in the following: IR) exhibits a confining potential of the form 
\be
V_{IR}(\zeta) = \lambda_A k_g \mu_g \,\zeta \, \left(1-\lambda_B \mu_g^2 \,\zeta^2\,e^{-\mu_g^2 \zeta^2}\right)\,,
\label{eq:Vconf_IR}
\ee
shown in Fig.\ref{fig:Vconf_IR}. The numerical values of the parameters are  as in Table \ref{tab:Vconf_IRparameters}.

The potential (\ref{eq:Vconf_IR}) belongs to the class of potentials obeying
$V(\zeta \to 0) = \alpha^2 \zeta$, and $V(\zeta \gg \mu_g^{-1})  =  \alpha^2 \zeta$, i.e. la class of potentials matching the linear wall both in the IR and UV regimes \cite{IR_isospectral}.

\begin{table}
\caption{Values of the parameters for $V_{IR}$ of Eq.(\ref{eq:Vconf_IR}). $\mu_g$, $\lambda_A$, $\lambda_B$ are from ref.\cite{IR_improvedN}. For $k_g$ see Eq.(\ref{eq:kg}) and discussion.}
\begin{center}
\begin{tabular}{llllllll}
\hline
$k_g$ (GeV) & \phantom{xx}& $\mu_g$ (GeV) &\phantom{xx}& $\lambda_A$ (u) &\phantom{xx}& $\lambda_B$ (u) \\ 
\hline
0.0089 && 0.473 && 3.93 && 16.58 \\
\hline
\end{tabular}
\end{center}
\label{tab:Vconf_IRparameters}
\end{table}%

Therefore the potential $V_{IR}$ must reduce to the linear soft-wall potential in the limiting case $\lambda_B=0$, and one has:
\be
\lambda_A k_g \mu_g = \alpha^2 = (0.41)^2\;{\rm GeV^2} \;\; \to k_g \approx 0.089\; {\rm GeV}\,,
\label{eq:kg}
\ee
parameters used in Fig.\ref{fig:Vconf_IR} and in the following.

The IR potential has been constructed to reproduce, with good accuracy, both the meson and the baryon masses. In particular it gives consistent predictions for the mass spectra of scalar, pseudoscalar, vector and axial-vector mesons, and both confinement and chiral symmetry breaking are well characterized \cite{IR_improvedM}.  In the case of baryons the parameters $\lambda_A$ and $\lambda_B$ are fixed by fitting the masses of the first low-lying baryons with even parity (including nucleon). The predicted masses for odd-parity baryons and high excited states of even-parity baryons are consistently reproduced \cite{IR_improvedN} by using the same values of the parameters.

Let us introduce the form (\ref{eq:Vconf_IR}) in the equations (\ref{eq:psiplus}), (\ref{eq:psiminus}), one gets
\bea
&& \left[-  {d^2 \over d\zeta^2} +{D^\pm\over \zeta^2}+ E^\pm \, \zeta^2 + F^\pm \,\zeta^2 \,e^{-\mu_g \zeta^2}+ G^\pm \,\zeta^4\, e^{-\mu_g \zeta^2} + \right. \nonumber \\
&& \left. + H^\pm \,\zeta^6\, e^{-2 \mu_g \zeta^2}+ I^\pm  \right] \phi^\pm_{\nu,IR} = \hat R^\pm \phi^\pm_{\nu,IR} = {\cal M}_{IR}^2 \phi^\pm_{\nu,IR}\,, \nonumber \\
\label{eq:psiminusplusVIR}
\eea
where ($A = \lambda_A k_g \mu_g$, $B = \lambda_B \mu_g^2$)

\be
\begin{array}{lcl}
 D^+ = - {(1-4\nu^2) / 4};     & &  D^- = - {[1-4(\nu+1)^2] /4}  \\ 
 & & \\
    E^+ = A^2; &  &E^- = A^2 \\
 & & \\
    F^+ = [-3 - (1+2 \nu)] A B;   &  &  F^- = [+3 - (1+2 \nu)] A B  \\ 
 & & \\
    G^+ = [-2A +2 \mu^2)] A B; &  & G^- = [-2A -2 \mu^2)] A B \\    
 & & \\
    H^+ = A^2 B^2 ;                   &  & H^- = A^2 B^2 \\
 & & \\  
    I^+  = 2 (1+\nu) A;                &  &  I^- =  2 \nu A \,.
\end{array}
\nonumber
\ee
A convenient technique to solve Eqs.(\ref{eq:psiminusplusVIR}) is an expansion on the basis of $\phi^\pm_{n l_\pm}$ of Eqs.(\ref{eq:sol-psipluslinear},\ref{eq:sol-psiminuslinear}), in this way one can keep all the already established properties of the solutions 
(\ref{eq:sol-psipluslinear},\ref{eq:sol-psiminuslinear}) within a linear combination of them.
Consequently 
\bea
\phi^\pm_{\nu,IR}(\zeta) = \sum_{n=0}^{n_{max}} a^\pm_{\nu, n} \, \phi^\pm_{n l_\pm}(\zeta)\,,
\label{eq:phi_exp}
\eea
where $\nu = 3$ and $l_+ = \nu$ and $l_- = \nu+1$ for the lowest three quark Fock state of the nucleon \cite{hFock}.
The natural parameter to be chosen to minimize ${\cal M}_{IR}^2$ looking for the ground state wave function (the nucleon) is the harmonic oscillator constant which has to be diversifyed in two components  
$\alpha \to \alpha^\pm$ in order to respect the essential property ${\cal M}_{IR}^+ = {\cal M}_{IR}^-$.
\be
\left. {\langle \phi^\pm_{\nu,IR}| \hat R^\pm | \phi^\pm_{\nu,IR}\rangle  \over \langle \phi^\pm_{\nu,IR}| \phi^\pm_{\nu,IR}\rangle}\right|_{minimum} \to ({\cal M^\pm}_{IR})^2 \to ({\cal M}_{IR})^2\,.
\label{eq:minimum0} \\
\ee
Of course the restricted Hilbert space used in solving the minimization will result in an upper bound for 
${\cal M}_{IR}^2$. However, as it will be more clear in the next Section, the convergence is rapid and one has to expect only few percent deviations. 

\subsection{\label{sec:numerical1}Numerical results}

The minimization procedure is performed in the two components $\phi^\pm_{\nu,IR}$ varying the parameter $\alpha^-$ and reaching the minimum value for $({\cal M}_{IR}^-)^2 = 2.61$ GeV$^2$ for $\alpha^- = 2.65$ fm$^{-1}$, with the corresponding $({\cal M}_{IR}^+)^2 = 2.61$ GeV$^2$ for $\alpha^+ = 2.35$ fm$^{-1}$, and involving 17 oscillator quanta ($n_{max} = 16$). The harmonic oscillator angular momentum quantum numbers $l_ - = \nu +1$ and $l_+ = \nu$ are fixed by the twist operator $\nu = 3$ for the lowest number of active quarks \cite{hFock} (of course $\sum_{n=0}^{n_{max}} (a^\pm_{n,l_\pm})^2 = 1$). In table \ref{tab:anl} the actual vales of the coefficients $a^\pm_{n,l_\pm}$. One can appreciate the rapid convergence. The basis is in fact the maximum numerical basis supported by the Matlab code used for the minimization, however it is evident that remaining within $n_{max}=10$ is a quite good approximation. The numerical calculations of the next Sections will make use of the restricted basis $n_{max}=10$.

\section{\label{sec:GPDs-SR}GPDs and sum rules at $\xi=0$}

\begin{figure}[tbp]
\centering\includegraphics[width=\columnwidth,clip=true,angle=0]{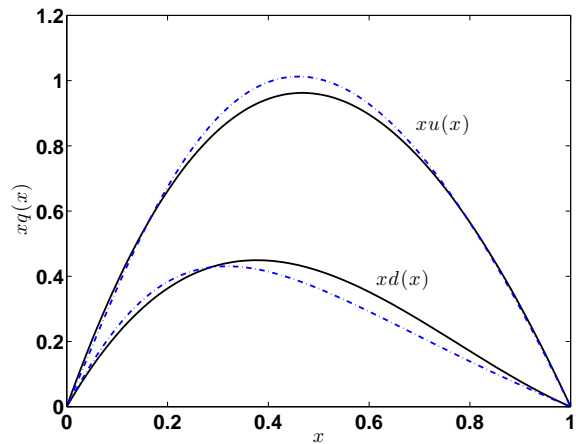}
\caption{\small (Color online) The distributions $x u(x)$ and $x d(x)$ as function of $x$. From Soft-Wall linear potential (solid lines) and from the IR-improved potential model (dot-dashed), when higher-Fock states are included (cfr. Section \ref{sec:higherFock}).The total momentum sum rule reads: $M_{u+d} = 0.92$ for the SW model, $M_{u+d}=0.91$ for the IR when higher Fock states are included.}
\label{fig:xq_SW_IR345}
\end{figure}

In order to introduce the explicit calculations of the GPDs, let us concentrate first on the chiral even (helicity conserving) distribution $H^q(x,\xi,Q^2,t)$ for partons  of $q$-flavor at the scale where one is assuming valid the calculation for the related amplitudes.  For example, the amplitude for deeply virtual Compton scattering where a virtual photon of momentum $q^\mu$ is exchanged by a lepton to a nucleon of momentum $P^\mu$ and a real photon of momentum $q'^\mu$ is produced (together with a recoiling nucleon $P'^\mu$).The space-like  virtuality is therefore $Q^2=-q^\mu q_\mu$ and it identifies the scale 
of the process.
The invariant momentum square is $t=-\Delta^2=(P'^\mu - P^\mu)^2$  and the skewedness $\xi$ encodes the change of the longitudinal nucleon momentum ($2 \xi = \Delta^+/\bar P^+$, with $2 \bar P^\mu = (P^\mu + P'^\mu)$). In the following the common notation of simply three variables $(x,\xi,t)$ instead of 
$(x,\xi,Q^2,t)$ is assumed\footnote{The chosen reference frame is symmetric and $q^\mu$ and the average moment $\bar P^\mu =(P^\mu+P'^\mu)/2$, are collinear (along the $z$ axis) and opposite in directions.}. In addition only the limit $\xi=0$ will be discussed and therefore one can remain in the $0 \leq x \leq 1$ region.

\subsection{\label{sec:nu3}Contribution of the valence quarks ($\nu=3$)}
\begin{table}
\caption{The numerical values of the coefficients $a_{\nu n}^\pm$ for the variational expansion (\ref{eq:phi_exp}) in the case of maximum h.o. quanta $n_{max}=16$ and $\nu=3$ ($l_+ = \nu = 3$, $l_-=\nu+1=4$). The h.o. constants are fixed by the minimization procedure at $\alpha^+ = 2.35$ fm$^{-1}$ and $\alpha^- = 2.65$ fm$^{-1}$.}
\begin{center}
\begin{tabular}{|r|r|r|}
\hline
     &              &   \\ 
     $n$ & $a_{(\nu=3), n}^+$ & $a_{(\nu=3), n}^-$   \\     
 \hline
     &              &   \\
0   & 0.9811  &  0.8749   \\
1   &-0.1872 &  -0.4423  \\
2   &0.0486    &  0.1834   \\
3   & -0.0071 & -0.0678  \\ 
4   &  0.0014  &  0.0233  \\
5   &  -0.0002 &-0.0076 \\
6   &3.6e-05    & 0.0024 \\
7   &-5.3e-06 & -7.3e-04 \\
8   & 8.3e-07  & 2.2e-04  \\
9   &-1.2e-07 & -6.5e-05 \\
10 & 1.8e-08  &1.9e-05 \\
11 &-2.7e-09  &-5.4e-06 \\
12 &  3.9e-10 & 1.5e-06\\
13 & -5.7e-11 & -4.3e-07\\
14 & 7.3e-12  & 1.2e-07  \\
15 &-5.7e-12 & -3.2e-08\\
16 &-6.4e-12 &  8.0e-09  \\
\hline
\end{tabular}
\end{center}
\label{tab:anl}
\end{table}%

The helicity conserving $H^q$ distributions, in the limit $t=0$ and $\xi=0$ reduce to ordinary parton distributions
\be
H^q(x,0,0) = q(x)\,,\label{eq:forward}
\ee
the unpolarized quark distribution of flavor $q$ and one has 
\be
\int dx \,H^q(x,0,0) = \int dx \, q(x) = N_q\,, \label{eq:Nq}
\ee
where $N_q$ fixes the number of valence quarks of flavor $q$ ($N_u =2$, $N_d = 1$).  The integral properties are therefore model independent and strongly constrain the helicity conserving distributions in any model and/or parametrization (the conditions on $N_q$ are satisfied within all the models presented). The second moment
\be
\sum_{q=u,d} \int dx\,x \,H^q(x,0,0) = \sum_{q=u,d} \int dx \,x\, q(x)  = {M_{u+d}}\,
\ee
is related to the momentum sum rule (cfr. Fig.\ref{fig:xq_SW_IR345}) and the models discussed differ: the LF-model is based on Light-Front wave functions and obeys $M_{u+d} = 1$ since the valence contribution is the only component at low momentum scale. The numerical calculations give: $M_{u+d} = 0.92$ for the SW, while $M_{u+d} = 0.91$ for the IR when higher Fock states are considered
(cfr. Section \ref{sec:higherFock}). In addition the first $t$-dependent moments of the GPDs are related to the nucleon elastic form factors \cite{GPDs_FF}, i.e.
\be
\int_{-1}^1 dx H^q(x,\xi,t) = F_1^q(\Delta^2)\,,\;\;\; \int_{-1}^1 E^q(x,\xi,t) = F_2^q(\Delta^2)\,, \label{eq:firstmoment}
\ee
where $F_1^q(\Delta^2)$ and $F_2^q(\Delta^2)$ are the contribution of quark $q$ to the Dirac and Pauli form factors. The property (\ref{eq:firstmoment}) does not depend on $\xi$ and it holds also in the present approach with $\xi=0$ and therefore $0 \leq x \leq 1$, (cfr refs.\cite{GPDs_FF,XJi_prize}), and one has:
\bea
F_1^p(\Delta^2) \!&= &\!\!\int_0^1 dx \left(+{2 \over 3} H^u_V(x,\xi\!=\!0, t) - {1 \over 3} H^d_V(x,\xi\!=\!0, t) \right) ,\nonumber \\
F_1^n(\Delta^2) \!&=& \!\! \int_0^1 dx \left(- {1 \over 3} H^u_V(x,\xi\!=\!0,t) + {2 \over 3} H^d_V(x,\xi\!=\!0,t) \right) ,\nonumber \\
F_2^p(\Delta^2) \! &=& \!\! \int_0^1 dx \left(+{2 \over 3} E^u_V(x,\xi\!=\!0,t) - {1 \over 3} E^d_V(x,\xi\!=\!0,t)\right) ,\nonumber \\
F_2^n(\Delta^2) \!&=&\!\! \int_0^1 dx \left(-{1 \over 3} E^u_V(x,\xi\!=\!0,t) + {2 \over 3} H^d_V(x,\xi\!=\!0,t)\right) \,,\nonumber \\
\eea
where $t = -\Delta^2$ and isospin symmetry has been assumed.
In terms of the holographic wave functions $\phi^\pm$ derived from $AdS/QCD$, the Dirac form factors for the nucleons in the present Soft-Wall  linear model are given by \cite{hep-ph/1407.8131,GPDs_HW,GPDs_SW,hFock}
\bea
F_1^p(\Delta^2) & = & \int d\zeta \, V^+(\Delta^2,\zeta)\, {|\phi^+_{\nu,IR}(\zeta)|^2 \over (\alpha^+ \zeta)^4}\,(N^+)^2 , \label{eq:F1pQ2} \\
F_1^n(\Delta^2) & = & -{1 \over 3} \int d\zeta\,  \left[V^+(\Delta^2,\zeta) {|\phi^+_{\nu,IR}(\zeta)|^2 \over (\alpha^+ \zeta)^4}\,(N^+)^2 + 
\right. \nonumber \\ 
&& \left. \!- V^-(\Delta^2,\zeta){|\phi^-_{\nu,IR}(\zeta)|^2 \over (\alpha^- \zeta)^4}\,(N^-)^2\right] , \label{eq:F1nQ2} \\
F_2^{p/n}(\Delta^2) & = & \kappa_{p/n} \times \nonumber \\
&\times & \!\! \int d\zeta {1 \over 2} \left[{\phi^-_{\nu,IR}(\zeta) V^-(\Delta^2,\zeta) \phi^+_{\nu,IR}(\zeta) \over (\alpha^- \zeta)^3}(N_\mp)^2 + \right. \nonumber \\ 
&&\left. \! +  {\phi^+_{\nu,IR}(\zeta) V^+(\Delta^2,\zeta) \phi^-_{\nu,IR}(\zeta) \over (\alpha^+ \zeta)^3}\,(N_\pm)^2 \right] \,;
\label{eq:F2pnQ2}
\eea
where $\kappa_{p/n}$ are the proton and neutron anomalous gyromagnetic factors respectively.
The kernels $V^\pm$ have a simple and analytic integral form \cite{VQz}:
\bea
&&V^\pm(\Delta^2,\zeta) = \int_0^1 dx\, F^\pm_x(\Delta^2,\zeta) = \label{eq:VQ2zetapm} \\
&& = \int_0^1 dx\, {({\alpha^\pm} \zeta)^2 \over (1-x)^2} \, x^{\Delta^2/[4 ({\alpha^\pm})^2]} e^{-({\alpha^\pm} \zeta)^2 x/(1-x)} . \nonumber 
\eea
The specific boundary condition $V^\pm(\Delta^2=0,\zeta) = 1$ imposes the normalizations:
\bea
&&\int d\zeta\, {|\phi^\pm_{\nu,IR}(\zeta)|^2 \over (\alpha^\pm \zeta)^4} \, (N^\pm)^2 = 1\,, \nonumber \\
&& \int d\zeta\, {\phi^-_{\nu,IR}(\zeta) \phi^+_{\nu,IR}(\zeta) \over (\alpha^- \zeta)^3}\,(N_\mp)^2 = 1\,, \nonumber \\ 
&& \int d\zeta\, {\phi^+_{\nu,IR}(\zeta) \phi^-_{\nu,IR}(\zeta) \over (\alpha^+ \zeta)^3}\,(N_\pm)^2 = 1 \,;
\eea
\\
and the results of the SW \cite{GPDs_SW} model are recovered in the limit 
\bea
&& \alpha^\pm \to  \alpha = 0.41\,\,{\rm GeV}, \nonumber \\
&& a_{n l_\pm}^\pm \to a_{n l_\pm} = 1, \nonumber \\
{\rm and\,\,therefore:} && \nonumber \\ 
&&\!\phi^\pm_{\nu,IR}(\zeta) \to \phi^\pm_{n l_\pm}(\zeta) \;\;{\rm of\,\,Eqs.(\ref{eq:sol-psipluslinear},\ref{eq:sol-psiminuslinear})}\nonumber \\
&& (N^+)^2 \to 2/(2/3!) = 6\,, \nonumber \\
&& (N^-)^2 \to 1/(2/4!) = 12 \,. \nonumber \\
&& (N_\pm)^2 \, \to (N_\mp)^2 =  6\,.
\label{eq:SWlimit}
\eea
The resulting expressions for the GPDs are
\bea
&+&{2 \over 3} H^u_V(x,\xi\!=\!0, -\Delta^2) - {1 \over 3} H^d_V(x,\xi\!=\!0, -\Delta^2) = \nonumber \\
&=& \int d\zeta \,F_x^+(\Delta^2,\zeta) \, {|\phi^+_{\nu,IR}(\zeta)|^2 \over (\alpha^+ \zeta)^4}\,(N^+)^2\,;
 \label{eq:Hu}
 \eea
 
 \bea
&-& {1 \over 3} H^u_V(x,\xi\!=\!0,-\Delta^2) + {2 \over 3} H^d_V(x,\xi\!=\!0,-\Delta^2) = \nonumber \\
&=& -{1 \over 3}\int d\zeta  \,\left[F_x^+(\Delta^2,\zeta) {|\phi^+_{\nu,IR}(\zeta)|^2 \over (\alpha^+ \zeta)^4}\,(N^+)^2 + \right. \nonumber \\
&& \left. \!- F_x^-(\Delta^2,\zeta){|\phi^-_{\nu,IR}(\zeta)|^2 \over (\alpha^- \zeta)^4}\,(N^-)^2\right]\,;
\label{eq:Hd}
\eea

\bea
&+&{2 \over 3} E^u_V(x,\xi\!=\!0, -\Delta^2) - {1 \over 3} E^d_V(x,\xi\!=\!0,-\Delta^2) = \nonumber \\
&=& \kappa_{p} \int d\zeta\,  {1 \over 2} \left[{\phi^-_{\nu,IR}(\zeta) F_x^-(\Delta^2,\zeta) \phi^+_{\nu,IR}(\zeta) \over (\alpha^- \zeta)^3}\,(N_\mp)^2  + \right. \nonumber \\ 
&&\left. \! +  {\phi^+_{\nu,IR}(\zeta) F_x^+(\Delta^2,\zeta) \phi^-_{\nu,IR}(\zeta) \over (\alpha^+ \zeta)^3}\,(N_\pm)^2 \right]\,; \label{eq:Eu} 
\eea

\bea
&-& {1 \over 3} E^u_V(x,\xi\!=\!0,-\Delta^2) + {2 \over 3} E^d_V(x,\xi\!=\!0,-\Delta^2) = \nonumber \\
&=& \kappa_{n} \int d\zeta\,  {1 \over 2} \left[{\phi^-_{\nu,IR}(\zeta) F_x^-(\Delta^2,\zeta) \phi^+_{\nu,IR}(\zeta) \over (\alpha^- \zeta)^3}\,(N_\mp)^2  + \right. \nonumber \\ 
&&\left. \! +  {\phi^+_{\nu,IR}(\zeta) F_x^+(\Delta^2,\zeta) \phi^-_{\nu,IR}(\zeta) \over (\alpha^+ \zeta)^3}\,(N_\pm)^2 \right]\,.
\label{eq:Ed}
\eea

\subsection{\label{sec:higherFock}Higher Fock states ($\nu=4$, $\nu=5$)}

The formalism developed in the previous Sections for the solution of the improved IR potential at the lowest twist ($\nu = 3$), can easily accommodate also higher Fock states in the wave functions opening the possibility of studying their effects on the Generalized Parton distributions even in presence of a modified potential. In particular additional gluons ($\nu = 4$) or a quark-antiquark pair ($\nu = 5$) as discussed in ref.\cite{hFock}. One obtains:

\bea
&&(\phi^\pm_{hF,IR}(\zeta))^2  = \sum _{\nu=3,4,5} c_\nu \,\,(\phi^\pm_{\nu,IR}(\zeta))^2\,,\nonumber \\
&& \phi^\pm_{hF,IR}(\zeta)  \, \phi^\mp_{hF,IR}(\zeta) = \sum _{\nu=3,4,5} c_\nu \,\,\phi^\pm_{\nu,IR}(\zeta) \, \phi^\mp_{\nu,IR}(\zeta)   \,, \nonumber \\
\label{eq:hFockwf} 
\eea
with $c_3 = 1.25$, $c_4 = 0.16$, and $c_5 = 1-c_3 - c_4 = - 0.41$.

The previous conditions and values are taken from ref.\cite{hFock} where they are established for the linear SW potential. However the criteria are rather general and directly related to experimental observables, their application also for the IR potential seems quite natural and it represents, in any case, a first sensible approximation. The minimization has to be repeated  for $\nu=4,5$ in analogy with the numerical analysis of Section \ref{sec:numerical1}. 
All the expressions derived in the previous Section \ref{sec:GPDs-SR} are generalized in a straightforward way replacing $(\phi^\pm_{\nu,IR}(\zeta))^2$ and $\phi^\pm_{\nu,IR}(\zeta) \, \phi^\mp_{\nu,IR}(\zeta) $
with the linear combinations (\ref{eq:hFockwf}). In order to comment in more detail, the generalization of the Eq.(\ref{eq:Hu})  is given as an example:
\bea
&+&{2 \over 3} H^u_V(x,\xi\!=\!0, -\Delta^2) - {1 \over 3} H^d_V(x,\xi\!=\!0, -\Delta^2) = \nonumber \\
&=& \int d\zeta \,F_x^+(\Delta^2,\zeta) \,\left[\sum_\nu c_\nu\, {|\phi^+_{\nu,IR}(\zeta)|^2 \over (\alpha^+ \zeta)^4}\,(N^+_\nu)^2\right]\,.
 \label{eq:Huagain}
 \eea
One has to notice that the normalization factors $N^2$ will depend on $\nu$ while the harmonic oscillator parameters $\alpha^\pm$ will not. In fact the baryon masses (fixed by the explicit form of the confining potential) will get their minimum values for the same $\alpha^\pm$, as it has been checked numerically. The generalization is straightforward. In the appendix the numerical details.

\begin{figure}[tbp]
\centering\includegraphics[width=\columnwidth,clip=true,angle=0]{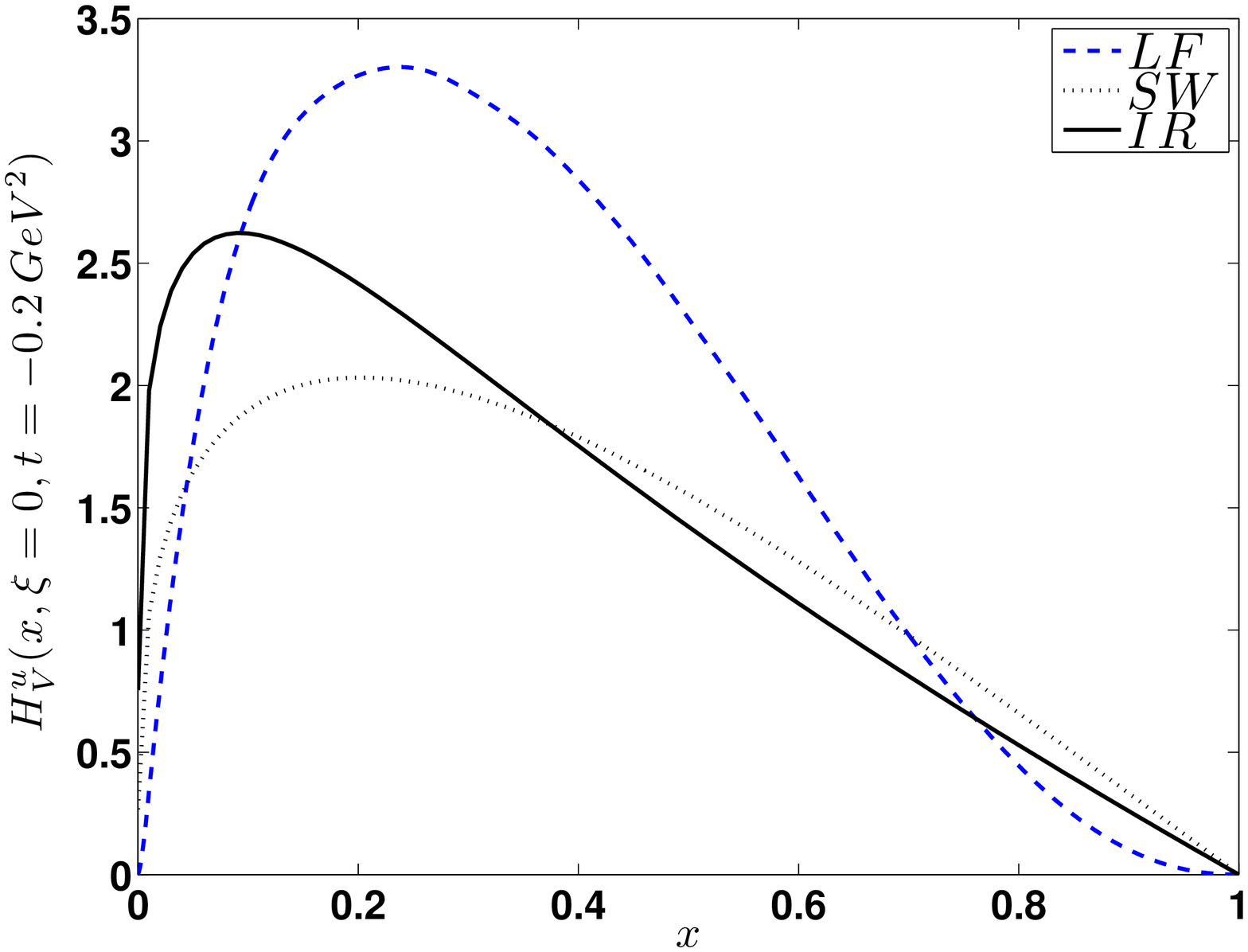}
\centering\includegraphics[width=\columnwidth,clip=true,angle=0]{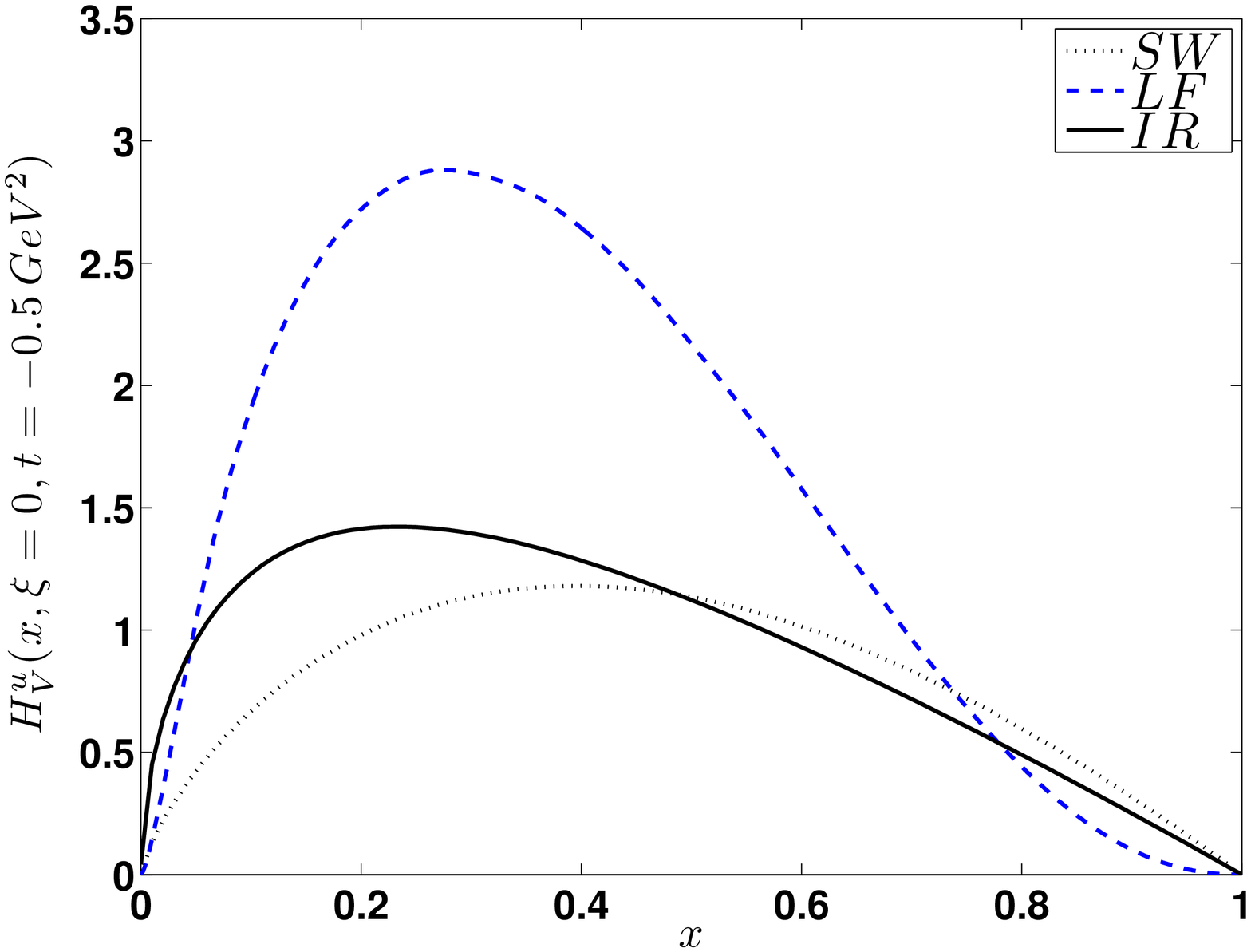}
\caption{\small (Color online) {\bf upper panel:} The results for $H_V^u(x,\xi=0,t=-0.2\,{\rm GeV}^2)$ predicted by the improved-IR model (continuous line) are compared with the same results for 
the corresponding SW model (dotted) and the LF model calculation of ref.\cite{BoffiPasquiniTraini1} (dashed). Only twist-3 contributions are included (cfr. Section \ref{sec:nu3}) and therefore the analysis is restricted to the valence sector.
{\bf lower panel:} As in the upper panel for $t=-0.5$ GeV$^2$. }
\label{fig:HuxQ2_02_05_IR_SW_LF}
\vspace{-1.0em}
\end{figure}

\begin{figure}[tbp]
\centering\includegraphics[width=\columnwidth,clip=true,angle=0]{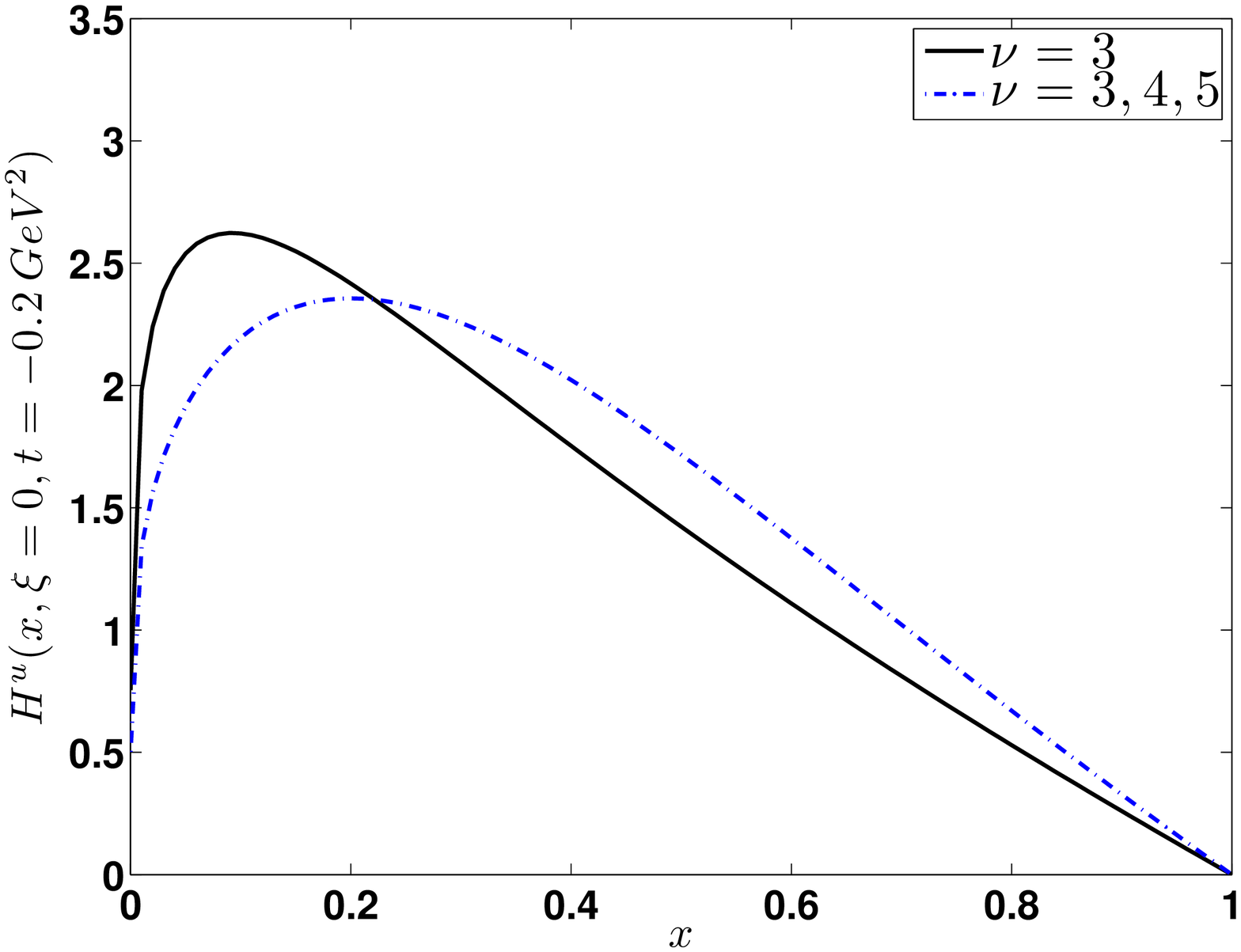}
\centering\includegraphics[width=\columnwidth,clip=true,angle=0]{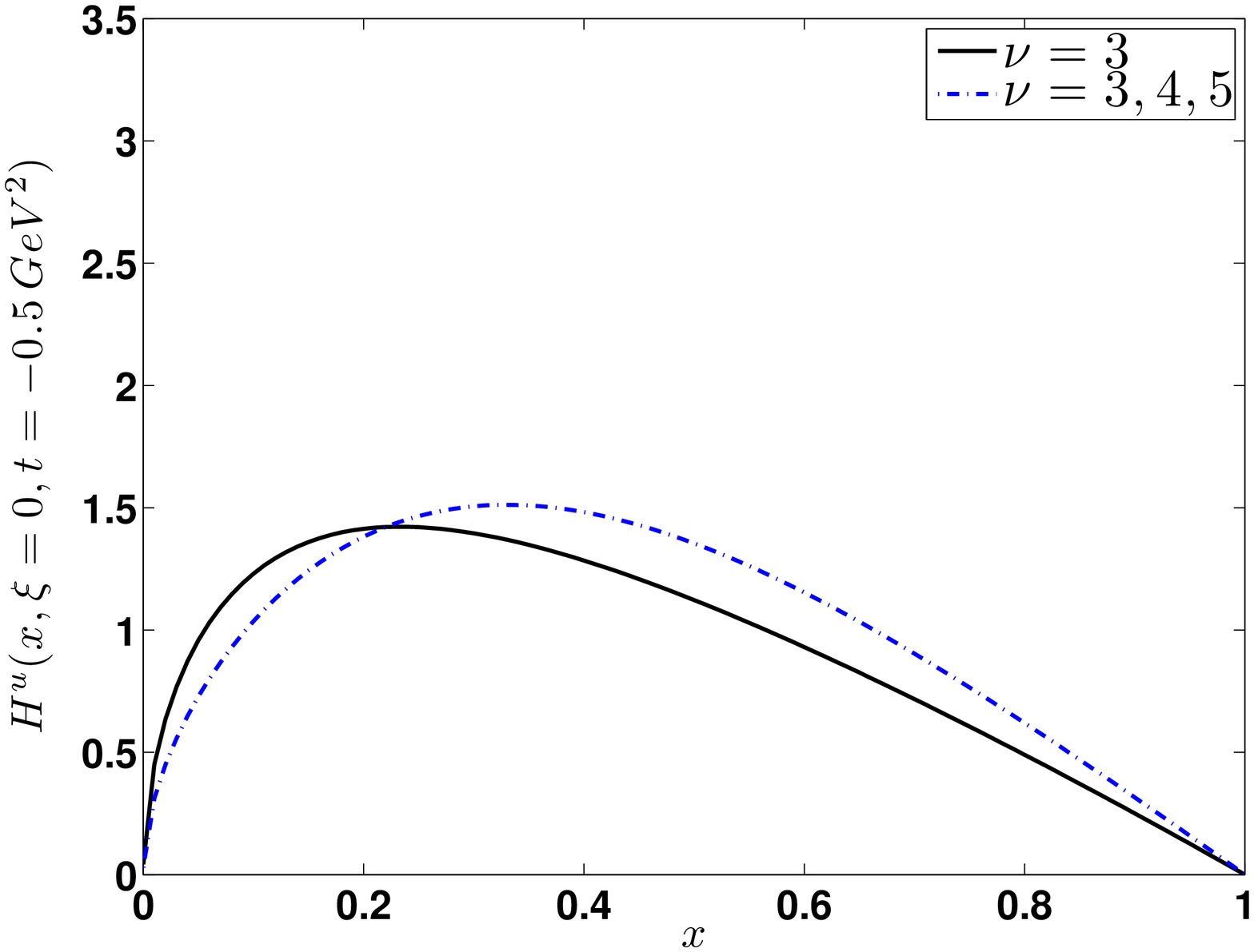}
\caption{\small (Color online) {\bf upper panel:} The results for $H_V^u(x,\xi=0,t=-0.2\,{\rm GeV}^2)$ predicted by the improved-IR model  and twist-3 contribution only (continuous line, the same results of Fig.\ref{fig:HuxQ2_02_05_IR_SW_LF}) are compared with the results obtained including higher Fock states (dot-dashed), namely $\nu=4,5$ (cfr. Section \ref{sec:higherFock}). 
{\bf lower panel:} As in the upper panel for $t=-0.5$ GeV$^2$. }
\label{fig:HuxQ2_02_05_IR_nu345}
\vspace{-1.0em}
\end{figure}

\section{\label{sec:results}GPDs and Confining Potentials: Results and comments}

\begin{figure}[tbp]
\centering\includegraphics[width=\columnwidth,clip=true,angle=0]{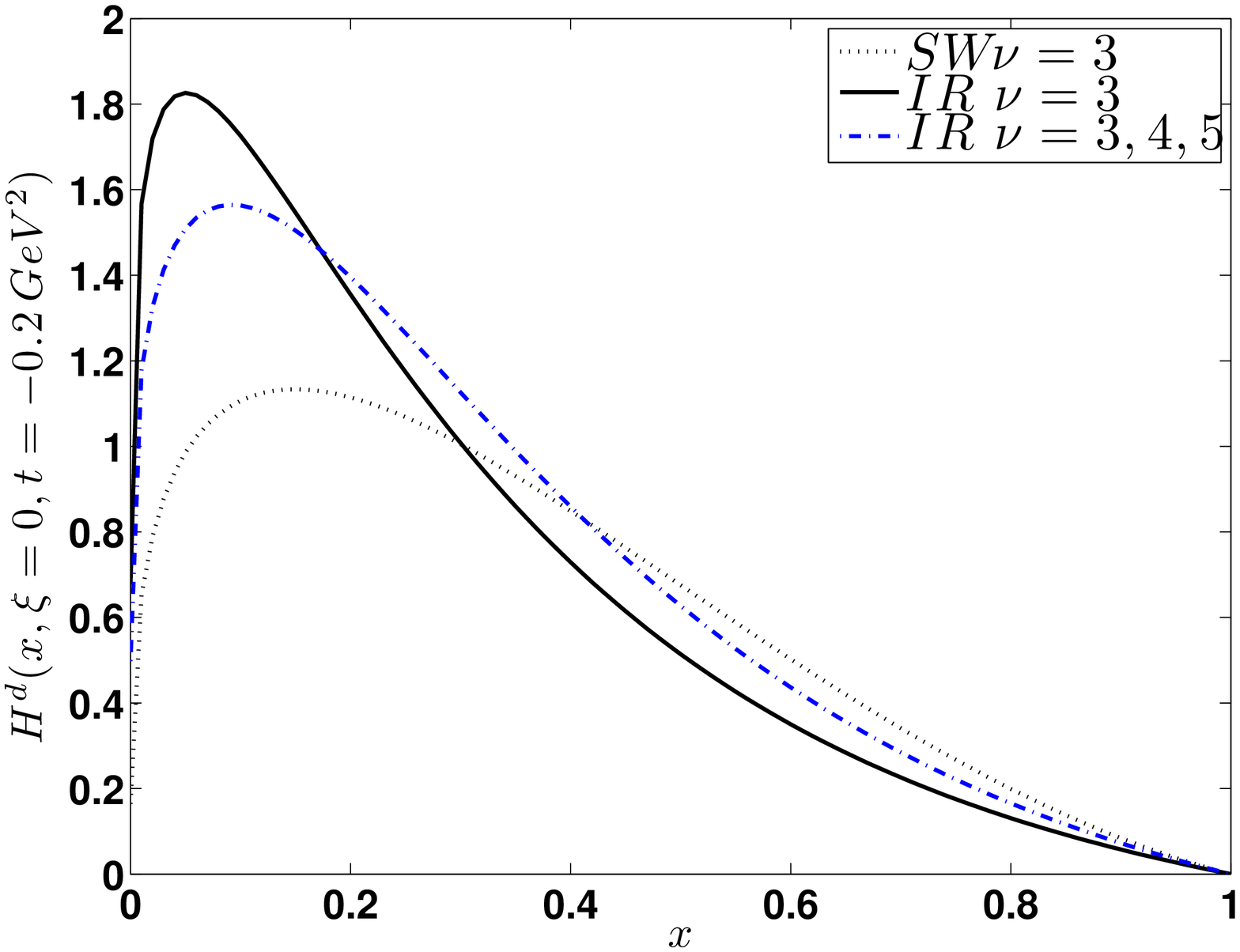}
\centering\includegraphics[width=\columnwidth,clip=true,angle=0]{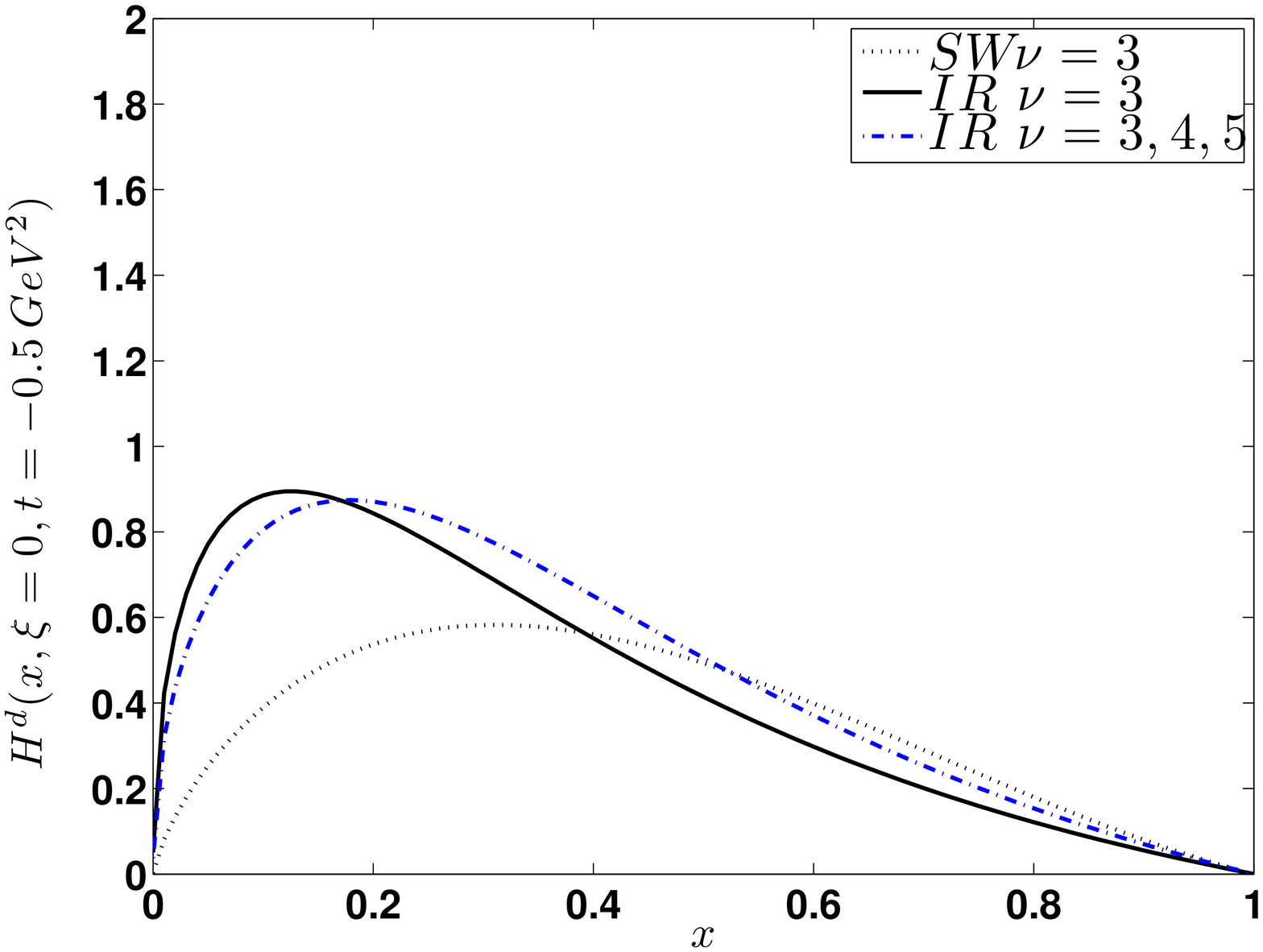}
\caption{\small (Color online) {\bf upper panel:} The results for $H^d(x,\xi=0,t=-0.2\,{\rm GeV}^2)$ predicted by the improved-IR model  and twist-3 contribution only (continuous line) are compared with the results obtained including higher Fock states (dot-dashed), namely $\nu=3,4,5$ (cfr. Section \ref{sec:higherFock}). The predictions of the corresponding SW model are also shown (dotted).
{\bf lower panel:} As in the upper panel for $t=-0.5$ GeV$^2$. }
\label{fig:HdxQ2_02_05_IR_nu345}
\vspace{-1.0em}
\end{figure}

\subsection{\label{sec:Hu}$H^u(x,\xi=0,t)$ and $H^d(x,\xi=0,t)$}

Results for $H^u(x,\xi=0,t)$ are shown in Figures \ref{fig:HuxQ2_02_05_IR_SW_LF} and \ref{fig:HuxQ2_02_05_IR_nu345}. In particular in Fig.\ref{fig:HuxQ2_02_05_IR_SW_LF}
the results for the valence components $H_V^u(x,\xi=0,t)$, i.e. the twist-3 contributions ($\nu=3$) are shown for both the SW model and the IR improved model. One could imagine that the change in the confining potential encodes just refinements producing only small effects on the observables. This comment is true from the point of view of the baryon spectra, however the modifications induced on the wave functions can show up in a more consistent way in appropriate observables. It is the philosophy of the present work and it is well illustrated in Fig.\ref{fig:HuxQ2_02_05_IR_SW_LF}: comparing the SW and the IR improved results one can appreciate the effects produced by the tuning of the confining potential (cfr. Fig.\ref{fig:Vconf_IR}). 
Analogous effects emerge in the analysis of the response of $d$-valence quarks 
(Fig.\ref{fig:HdxQ2_02_05_IR_nu345}). In that case the effects of the IR improved potential seem to be even more evident in the low-$x$ region and for both $t=-0.2$ GeV$^2$ and $t=-0.5$ GeV$^2$.

The $t$-dependence of the $H$-GPDs can be appreciated comparing the upper and lower panels of 
Figs.\ref{fig:HuxQ2_02_05_IR_SW_LF} and \ref{fig:HdxQ2_02_05_IR_nu345} where the responses are shown for two different values of the momentum $t=-0.2$ GeV$^2$ and $t=-0.5$ GeV$^2$. In particular in Fig.\ref{fig:HuxQ2_02_05_IR_SW_LF} the results of the present AdS/QCD approach are compared with an investigation (cfr. ref.\cite{BoffiPasquiniTraini1})  which makes use of a Light-Front relativistic quark model developed in ref.\cite{FaccioliTrainiVento} and based on a $q$-$q$-potential with a linear plus a Coulomb-like component: $V = -{\tau \over r} + \kappa_l \, r$ . 
The predictions of the two approaches look rather different. The constrains due to conformal symmetry breaking imposed by the AdS/QCD approach seems to reduce  the response considerably (and in the whole $x$-range) changing, at the same time, their $t$-dependence in a relevant way.

Figures \ref{fig:HuxQ2_02_05_IR_nu345} and \ref{fig:HdxQ2_02_05_IR_nu345} are devoted to the investigation of the higher Fock states effects. Within the IR-improved potential the $\nu=3$ and $\nu=3,4,5$ responses are shown and compared. {The effects of  higher-Fock states is rather weak, but one has to keep in mind the limited validity of the contribution for $\xi=0$, the only component here discussed. The role of quark-antiquark and gluon components should show up in a more consistent way in the $\xi$-dependence of the response \cite{DiehlPR2003,GPDs_NonP-P,GPDs_cloud}. It would be particularly interesting, in view of the next generation of experiments, to add explicitly such components together with the appropriate perturbative QCD evolution. Work in this direction is in progress.}

The comparison with experiments seems also particularly interesting from the point of view of the $t$-dependence of the responses. Often such a dependence is taken following the fall off of the nucleon form factors. Modeling GPDs does not confirm that hypothesis and the results of ref.\cite{BoffiPasquiniTraini1} already questioned such a $t$-dependence. The results of the AdS/QCD approach show an even stronger $t$-dependence, a peculiarity which should be explicitly investigated in future experiments.

\begin{figure}[tbp]
\centering\includegraphics[width=\columnwidth,clip=true,angle=0]{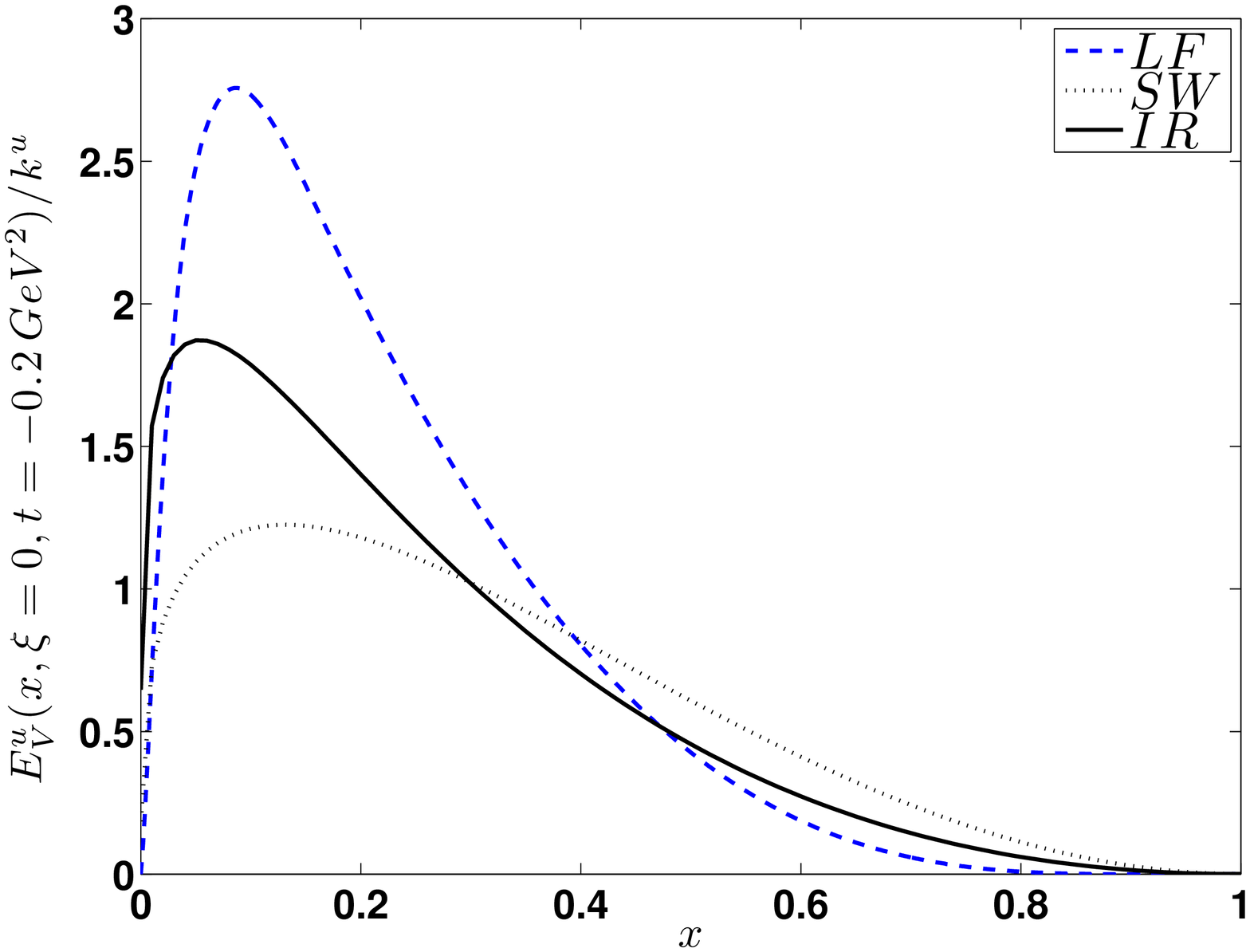}
\centering\includegraphics[width=\columnwidth,clip=true,angle=0]{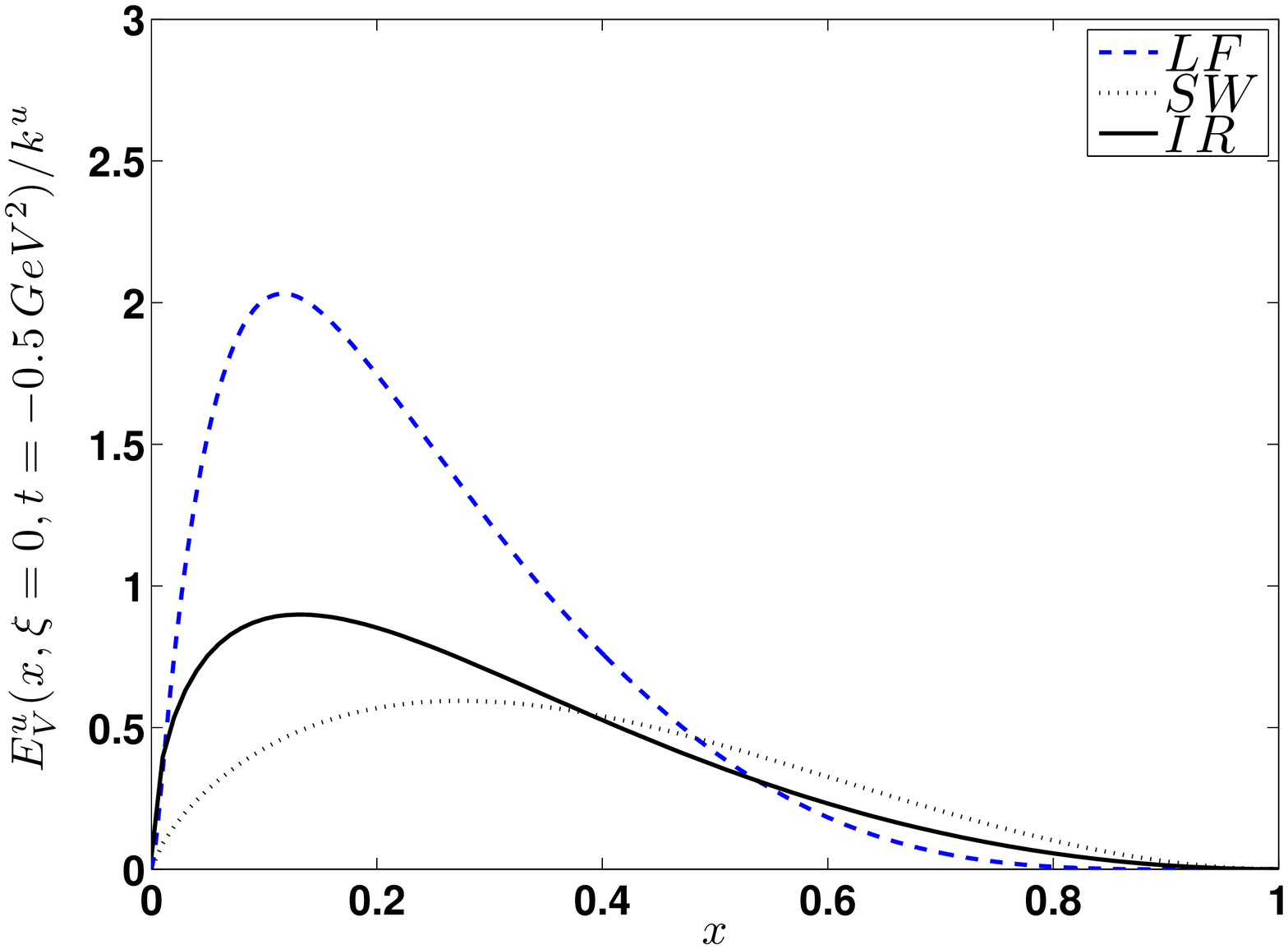}
\caption{\small (Color online) {\bf upper panel:} The results for $E_V^u(x,\xi=0,t=-0.2\,{\rm GeV}^2)$ predicted by the improved-IR model (continuous line) are compared with the same results for a
the corresponding SW model (dotted) and the LF model calculation of ref.\cite{BoffiPasquiniTraini1} (dashed). Only twist-3 contributions are included (cfr. Section \ref{sec:nu3}) and therefore the analysis is restricted to the valence sector.
{\bf lower panel:} As in the upper panel for $t=-0.5$ GeV$^2$. }
\label{fig:EuxQ2_02_05_IR_SW_LF}
\vspace{-1.0em}
\end{figure}

\subsection{\label{sec:Eud}$E^u(x,\xi=0,t)$ and $E^d(x,\xi=0,t)$}

\begin{figure}[tbp]
\centering\includegraphics[width=\columnwidth,clip=true,angle=0]{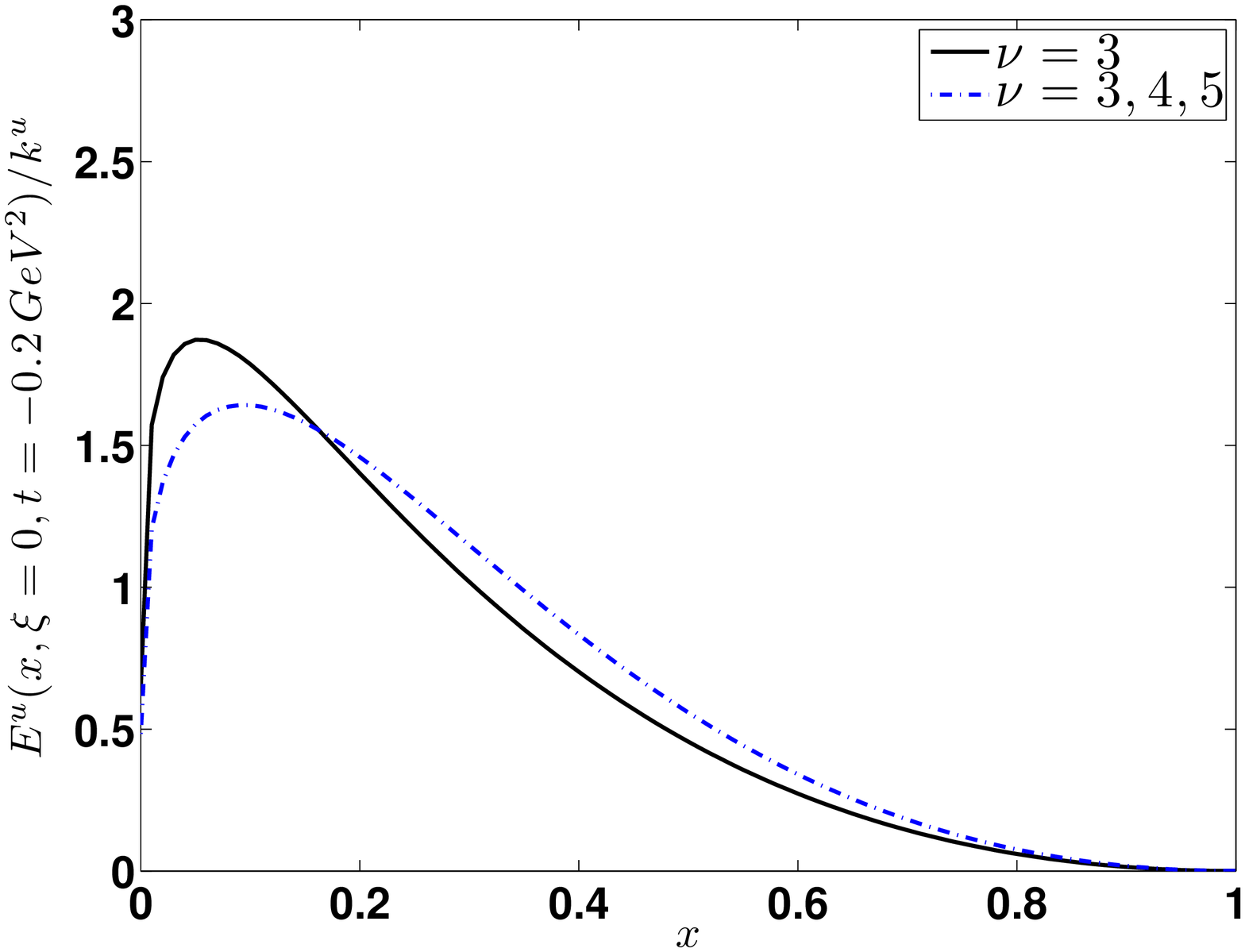}
\centering\includegraphics[width=\columnwidth,clip=true,angle=0]{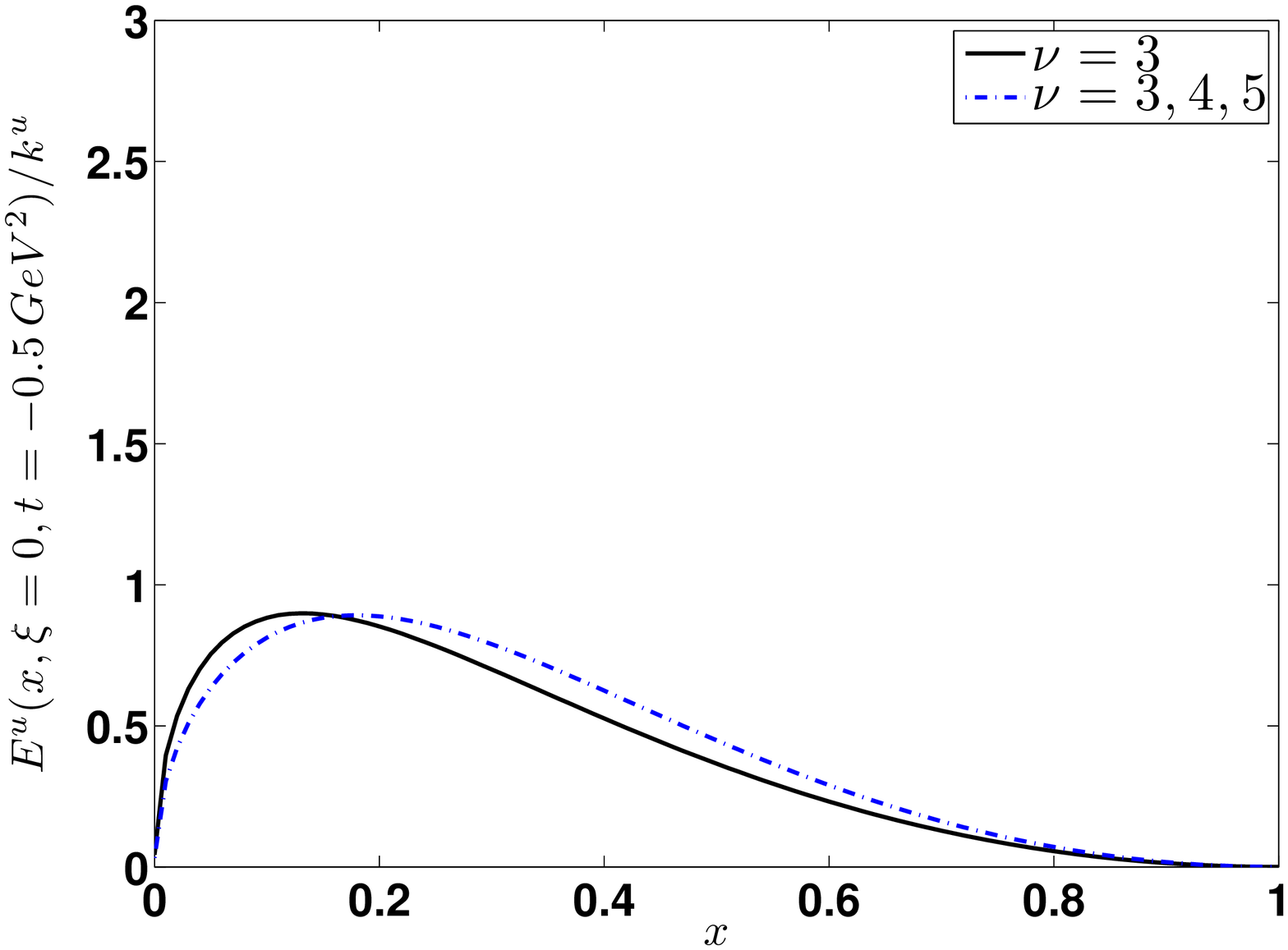}
\caption{\small (Color online) {\bf upper panel:} The results for $E_V^u(x,\xi=0,t=-0.2\,{\rm GeV}^2)$ predicted by the improved-IR model  and twist-3 contribution only (continuous line, the same results of Fig.\ref{fig:EuxQ2_02_05_IR_SW_LF}) are compared with the results obtained including higher Fock states (dot-dashed), namely $\nu=3,4,5$, cfr. Section \ref{sec:higherFock}). 
{\bf lower panel:} As in the upper panel for $t=-0.5$ GeV$^2$. }
\label{fig:EuxQ2_02_05_IR_nu345}
\vspace{-1.0em}
\end{figure}

\begin{figure}[tbp]
\centering\includegraphics[width=\columnwidth,clip=true,angle=0]{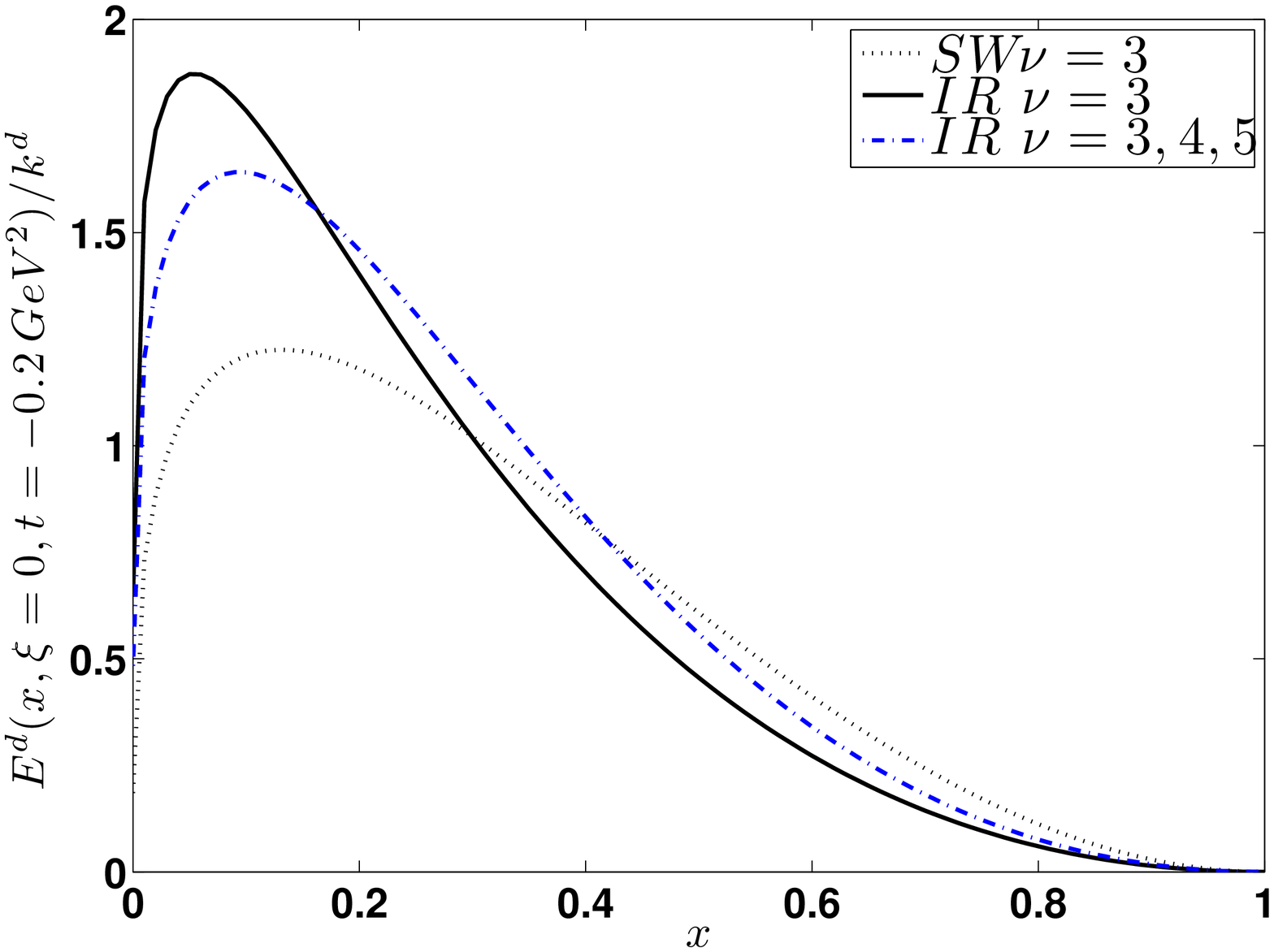}
\centering\includegraphics[width=\columnwidth,clip=true,angle=0]{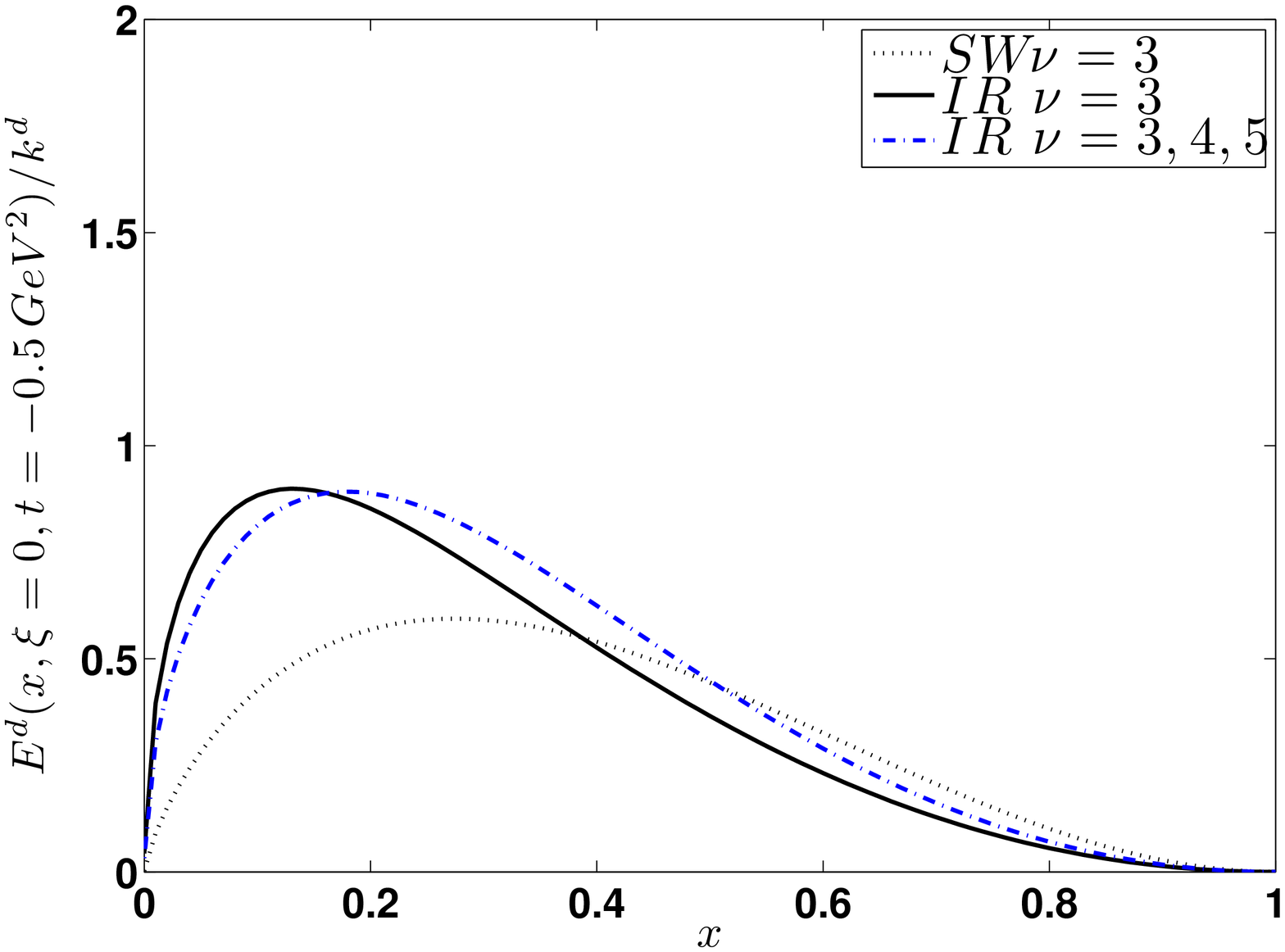}
\caption{\small (Color online) {\bf upper panel:} The results for $E^d(x,\xi=0,t=-0.2\,{\rm GeV}^2)$ predicted by the improved-IR model  and twist-3 contribution only (continuous line) are compared with the results obtained including higher Fock states (dot-dashed), namely $\nu=3,4,5$, cfr. Section \ref{sec:higherFock}). The predictions of the corresponding SW model are also shown (dotted).
{\bf lower panel:} As in the upper panel for $t=-0.5$ GeV$^2$. }
\label{fig:EdxQ2_02_05_IR_nu345}
\vspace{-1.0em}
\end{figure}

The integral properties  of the helicity non-conserving responses $E^q$ are more model dependent:
\bea
\int dx \, E^q(x,0,0) = \kappa_q\,,
\eea
where $\kappa^q$ is the anomalous magnetic moment. Experimentally $\kappa^u = 2\kappa^p+\kappa^n = 1.67$ and $\kappa^d = 2 \kappa^n+\kappa^p = -2.03$. From a theoretical point of view the calculation of  $\kappa^q$ is affected by the dynamical hypothesis of the approach used. In particular for the LF-approach of ref.\cite{BoffiPasquiniTraini1} one has $\kappa^u = 1.02$ and $\kappa^d = -0.74$\footnote{More explicitly, ref.\cite{BoffiPasquiniTraini1} investigates two L-F quark models: i) a Hypercentral potential which includes linear and Coulombian interactions and which is $SU(6)$ symmetric; ii) a model with Goldstone Boson Exchange (GBE) \cite{GBE1} which breaks $SU(6)$. Despite the fact that $\kappa_{p/n}$  are in principle  sensitive to $SU(6)$ breaking effects, the vales of the two models do not differ that much. For details cfr. ref.\cite{BoffiPasquiniTraini1}.}. On the contrary the AdS/QCD wave functions are normalized at the experimental values. To make the comparison more meaningful Figs.\ref{fig:EuxQ2_02_05_IR_SW_LF}, \ref{fig:EuxQ2_02_05_IR_nu345} and \ref{fig:EdxQ2_02_05_IR_nu345} show the ratios $E^q(x,\xi,t)/\kappa^q$. They are in continuity with Figs.\ref{fig:HuxQ2_02_05_IR_SW_LF}, \ref{fig:HuxQ2_02_05_IR_nu345} and \ref{fig:HdxQ2_02_05_IR_nu345} for the $H^q$ responses.

Also for the $E^q$ distributions the SW and the IR-improved potentials predict significantly different results as far as their $x$-dependence is concerning. The comparison with the LF-approach shows also an important difference in $t$- dependence between the LF and the AdS/QCD approaches. The inclusion of higher Fock states is illustrated in Figs.\ref{fig:EuxQ2_02_05_IR_nu345} and \ref{fig:EdxQ2_02_05_IR_nu345}.

\subsubsection{\label{sec:alpha}The $\alpha$ parameter: a critical analysis}

\begin{figure}[tbp]
\centering\includegraphics[width=\columnwidth,clip=true,angle=0]{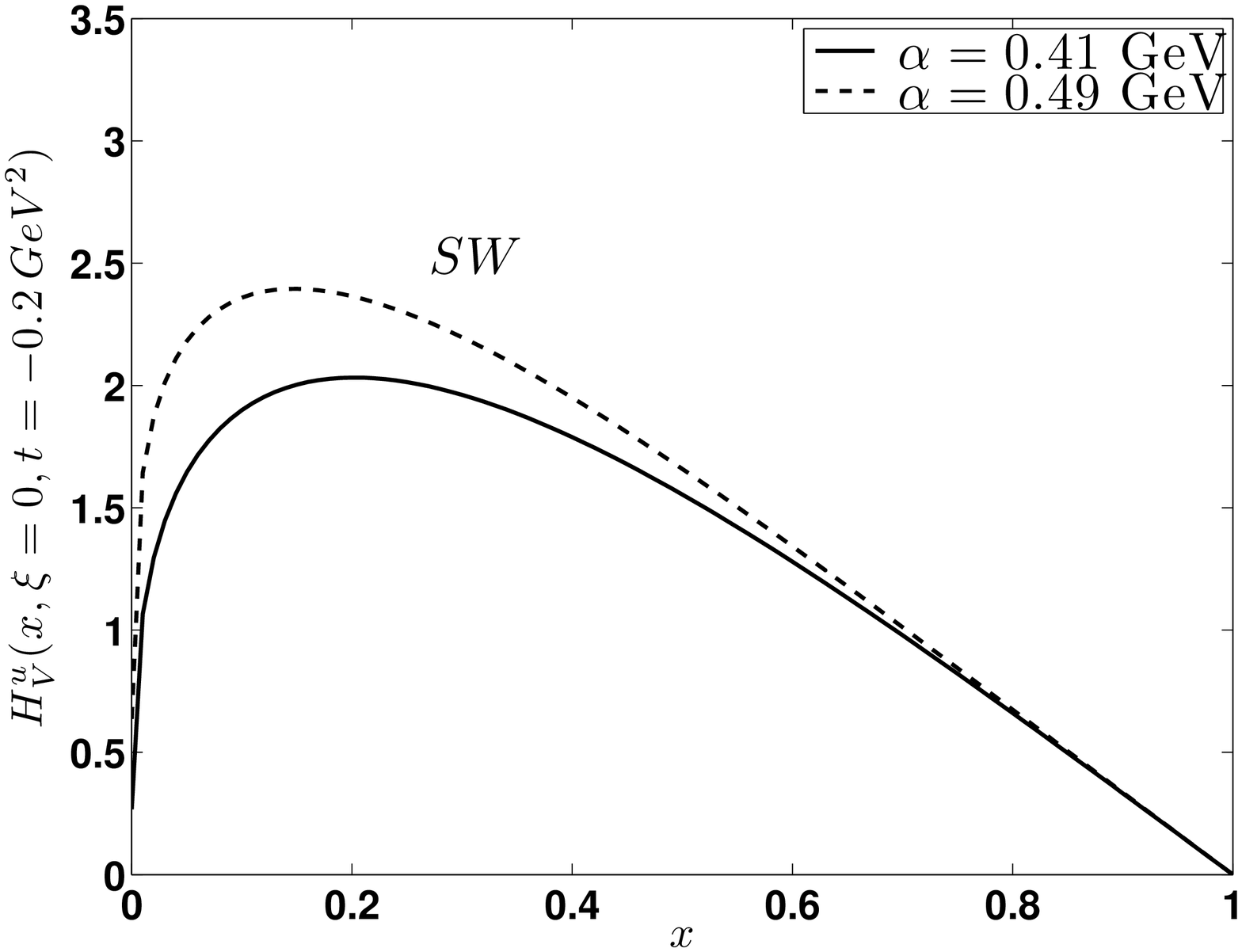}
\centering\includegraphics[width=\columnwidth,clip=true,angle=0]{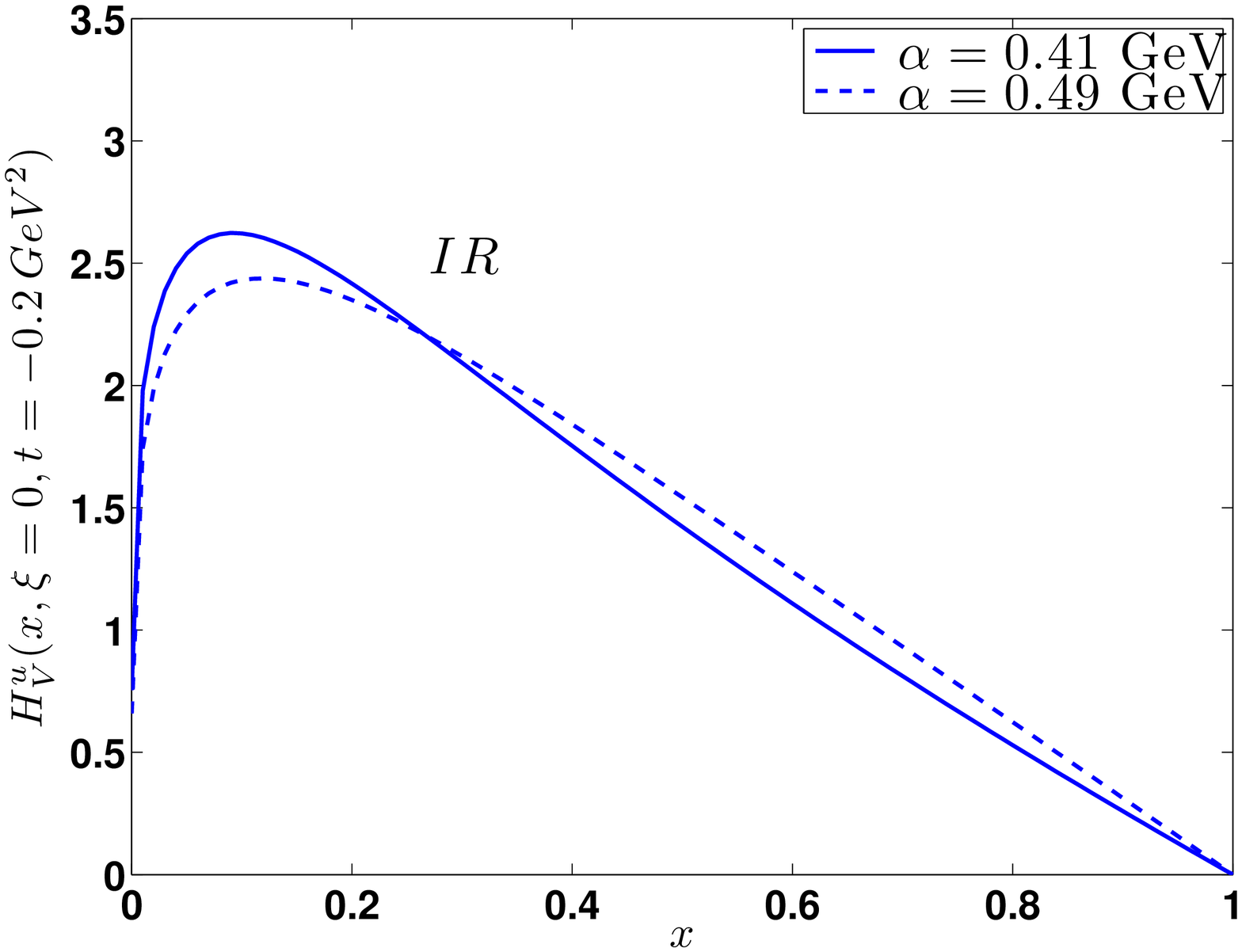}
\caption{\small (Color online) {\bf upper panel:} The sensitivity of  $H_V^u(x,\xi=0,t=-0.2\,{\rm GeV}^2)$  to the variation of the $\alpha$ parameter in the case of SW predictions. 
{\bf lower panel:} As in the upper panel for $IR$ improved potential. See text.}
\label{fig:HuxQ2_02_IRSW_alpha}
\vspace{-1.0em}
\end{figure}

Before discussing some application of the GPDs in \\ AdS/QCD,
a critical analysis of the parameter $\alpha$ characterizing the SW potential (cfr. Section \ref{sec:SW-linear}) could help in fixing the precision one can expect in the present, and analogous, investigations.
To this end it is convenient to write explicitly the GPDs within the SW approach, as they result from
Eqs.(\ref{eq:Hu}) - (\ref{eq:Ed}) in the SW-limit of Eq.(\ref{eq:SWlimit}) \footnote{Results for the helicity independent components will be discussed, they are illustrative also for the results of the helicity dependent component.}: 
\bea
H^u_V(x,\xi=0,\mu_0^2,t) & = &  u_V(x,\mu_0^2) \,x^{-{t\over 4 
\alpha^2}}, 
\nonumber \\
H^d_V(x,\xi=0,\mu_0^2,t) & = & d_V(x,\mu_0^2)\, 
x^{-{t \over 4 
\alpha^2}}~.
\label{eq:HudSW} 
\eea
(with $ t = -\Delta^2$). It is evident from Eq.(\ref{eq:HudSW}),  (and analogous expressions can be written for the helicity dependent components)  that, for $t=0$, the parameter $\alpha$ does not affect the $x$-dependence of  $H^q$, it influences its $t$-dependence. Such a conclusion has the relevant consequence that the differences one can see in Fig.\ref{fig:xq_SW_IR345} are $\alpha$-independent. For $t < 0$ the effects are more complicated correlating in a critical way the $x$ and $t$-dependence\footnote{These correlations can have important physical consequences in double (or multiple) parton scattering (e.g. ref.\cite{Traini_etal_PLB2017}).} and the choice of the $\alpha$-parameter appears to be critical. 
The discussion of the SW spectrum for baryons shows that the masses obey the Regge behavior  and $\alpha \approx 0.5$ GeV is needed to reproduce the nucleon spectrum (cfr. Section
\ref{sec:SW-linear}). In the literature the values $\alpha = 0.49$ GeV and $\alpha = 0.51$ are considered the best choices to reproduce, within the holographic AdS/QCD,  the nucleon and the $\Delta$ spectra respectively \cite{hep-ph/1407.8131}. The freedom in the choice of $\alpha$ is related to the nature of the AdS/QCD approach and the actual value is fixed following physical constraints like the nucleon and the $\Delta$ masses. In the study of the nucleon electromagnetic form factors, $\alpha$ is fixed in order to reproduce their momentum transfer behavior and, to this end, it has been chosen \cite{GPDs_SW2} $\alpha = 0.4066 \approx 0.41$ GeV. It is physically sensible to remain within this choice in order to study GPDs. However just to give a flavor of the $\alpha$ dependence of the present investigation in Fig.\ref{fig:HuxQ2_02_IRSW_alpha} the sensitivity of the helicity independent GPDs to the $\alpha$'s values is shown for both the SW and IR improved potential. The values chosen are: i) the choice made in the previous Sections and related to the electromagnetic form factors, $\alpha = 0.41$ GeV; ii) the value from the best fit of the nucleon masses, $\alpha = 0.49$ GeV. The variations shown in Fig.\ref{fig:HuxQ2_02_IRSW_alpha} could represent an upper bound to the absolute theoretical error. 
However,  one cannot consider the range of the results shown in the figure as genuine theoretical error bars; in fact the value $\alpha = 0.41$ GeV is well constrained to be associated to the electromagnetic interactions as described within AdS/QCD. One has to keep in mind, indeed, that  $\alpha$ is the parameter that appears in the dilaton definition used to break conformal invariance in AdS and it affects all fields considered in the model, including the vector massless field which allows the calculation of form factors (and GPDs), cfr. Eq.(\ref{eq:VQ2zetapm}). The same parameter appears, in the case of the nucleon, in the Soft-Wall potential, in the holographic coordinate $V(z) = \alpha^2 \, z$.  Consistency is mandatory and the $\alpha$ value has to be fixed by physical constrains connected with the vector massless field dual to the electromagnetic field and the form factors appear a natural choice. 
\\

In concluding the present Section \ref{sec:results}, a general comment can be added in order to justify the large differences one can see in the IR versus SW potential predictions as well as in the comparison with the LF model. The quite different behavior of the potentials at intermediate vales of $z \approx 0.5$ fm (see Fig.\ref{fig:Vconf_IR}), introduce relevant differences in the high-momentum components of the wave corresponding functions and, consequently, on the $H$-distributions. In particular, if  the behavior of the IR-potential is extrapolated to small distances ($< 0.5$ fm) to match Coulomb tail like in the case the LF model, the enhancement at small- and intermediate-$x$ values is emphasized as it emerges, for instance, from Figs.(\ref{fig:HuxQ2_02_05_IR_SW_LF}) and (\ref{fig:EuxQ2_02_05_IR_SW_LF}). The responses of the LF model (in the region $0 \leq x \leq 0.3$) are larger than the IR potential ones; the IR responses are, in turn, larger than the SW model distributions: a coherent behavior. On the contrary, because of the sum rule constrains, the responses in the large-$x$ region follow an inverse behavior.{
\section{\label{sec:modeling} Modeling the $\xi$-dependence with double distributions}

In the present Section the results obtained at $\xi=0$, are generalized to the whole $\xi$ domain by means of a double distribution approach developed by Radyushkin in ref.\cite{two_component1999}. The approach  involves a given profile function and the forward parton distribution as evaluated in the previous Sections (or in a generic model). 
In order to be specific let us concentrate on the chiral even (helicity conserving) distributions
$H^q(x,\xi,Q^2,t)$. One can introduce NonSinglet  (valence) and Singlet quark distributions:
\bea
H^{\rm NS}(x,\xi,t) & \equiv& \sum_q \left[H^q(x,\xi,t) + H^q(-x,\xi,t)\right] \nonumber \\
& = & + H^{\rm NS}(-x,\xi,t), \label{eq:sym}\\
H^{\rm S}(x,\xi,t) & \equiv& \sum_q \left[H^q(x,\xi,t) - H^q(-x,\xi,t)\right], \nonumber \\
& = & - H^{\rm S}(-x,\xi,t). \label{eq:antisym}
\eea
The analogous distribution for gluons is symmetric in $x$,
\be
H^g(x,\xi,t) = H^g(-x,\xi,t), \label{eq:symg}
\ee
with
\be
H^g(x,\xi=0,t=0) = x g(x), \;\;\; x> 0.
\ee 
Once again, the $Q^2$ dependence has been omitted following the common simplified notation, it will be discussed in Section \ref{sec:Q02} when the hadronic scale $Q_0^2$ will be introduced. Due to the polynomiality property \cite{polynomiality1998} the symmetry characters,  (\ref{eq:sym}), (\ref{eq:antisym}) and (\ref{eq:symg}),  hold also under $\xi \to - \xi$. The Singlet and gluon components mix under evolution, while the NonSinglet distribution evolve independently.

The $t$-independent part can be parametrized by a two component form \cite{two_component1999}
\be
H^q(x,\xi) = H^q_{DD}(x,\xi) + \theta(\xi - |x|)\,D^q\left(x \over \xi\right),
\label{eq:DD1}
\ee
with
\be
H^q_{DD}(x,\xi) = \int_{-1}^{+1} d\beta \int_{-1 + |\beta|}^{1-|\beta|} d\alpha \, \delta(x-\beta-\alpha \xi)\,F^q(\beta,\alpha), \label{eq:DD2}
\ee
and $H^q(x,\xi) \equiv H^q(x,\xi,t=0)$. 

The $D^q$ contribution in Eq.(\ref{eq:DD1}) is defined in the region $|x| \leq \xi$ and therefore does not contribute in the forward limit. The D-term contributes to the Singlet-quark and gluon distributions and does not contribute to NonSinglet components. Its effect under evolution is restricted at the level of few percent \cite{GPDs_EV} and it will be disregarded in the following.

Following Radyushkin the DD terms entering Eq.(\ref{eq:DD2}) are written as 
\be
F^q(\beta,\alpha) = h(\beta,\alpha) \, H^q(\beta,0,0),
\ee
where $H^q(\beta,0,0) = q(\beta)$ (cfr. Eq.(\ref{eq:forward})) and the profile function is parametrized
as \cite{DD_Radyushkin2001}
\be
h(\beta,\alpha) = {\Gamma(2 b + 2) \over 2^{2b+1}\Gamma^2(b+1)} {\left[ (1-|\beta|)^2 - \alpha^2 \right]^b \over (1-|\beta|)^{2b+1}}.
\ee
The parameter $b$ fixes the width of the profile function $h(\beta,\alpha)$ and the strength of the $\xi$-dependence. In principle it could be used (within the double distribution approach) as a fit parameter in the extraction of GPDs from hard electro-production observables. The favored choice is
$b_{\rm NS} = b_{\rm S} = 1$ (producing a maximum skewedness) and $b_{gluon} = 2$ \cite{DD_Radyushkin2001,GPDs_NonP-P}. In the limiting case $b \to \infty$, $h(\beta,\alpha) \to \delta(\alpha) h(\beta)$ and $H^q(x,\xi) \to H^q(x,\xi=0)$. The explicit evaluation of  $H^q(x,\xi)$ in Eq.(\ref{eq:DD1})  makes use of the results of the previous Sections within the holographic AdS/QCD approach.

\subsection{\label{sec:Q02}Results at low momentum scale: \\ the Soft-Wall model}

\begin{figure}[tbp]
\centering\includegraphics[width=\columnwidth,clip=true,angle=0]{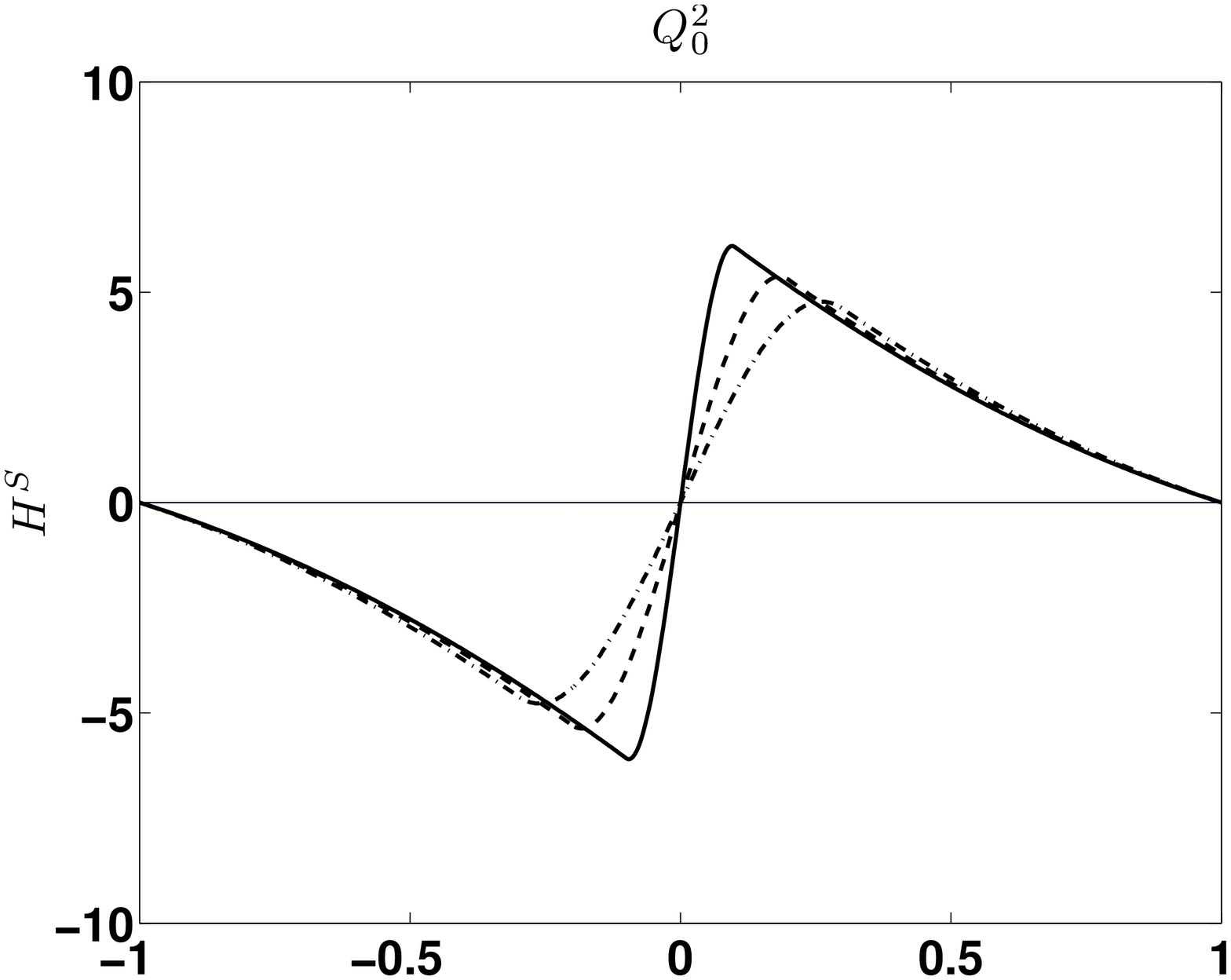}
\centering\includegraphics[width=\columnwidth,clip=true,angle=0]{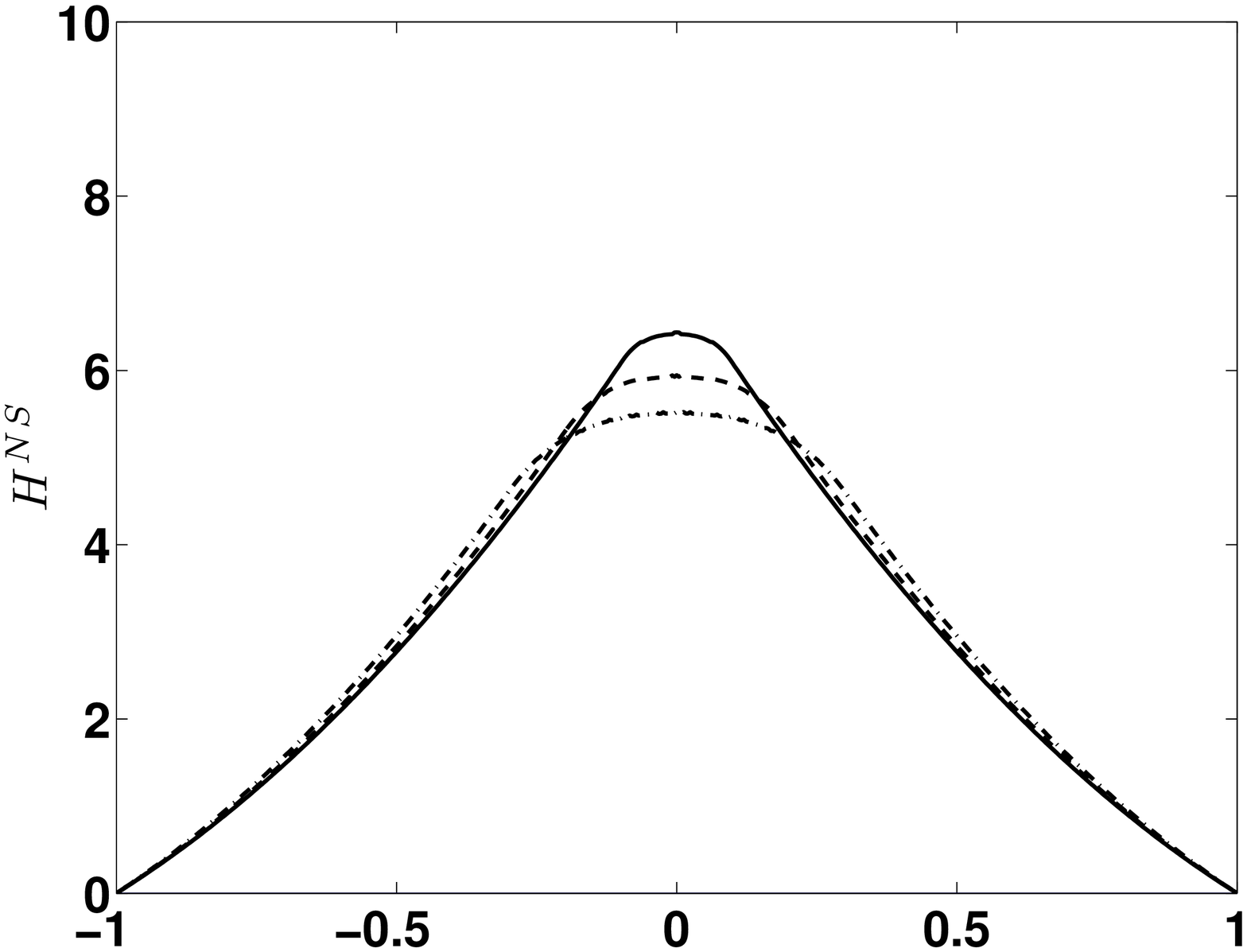}
\centering\includegraphics[width=\columnwidth,clip=true,angle=0]{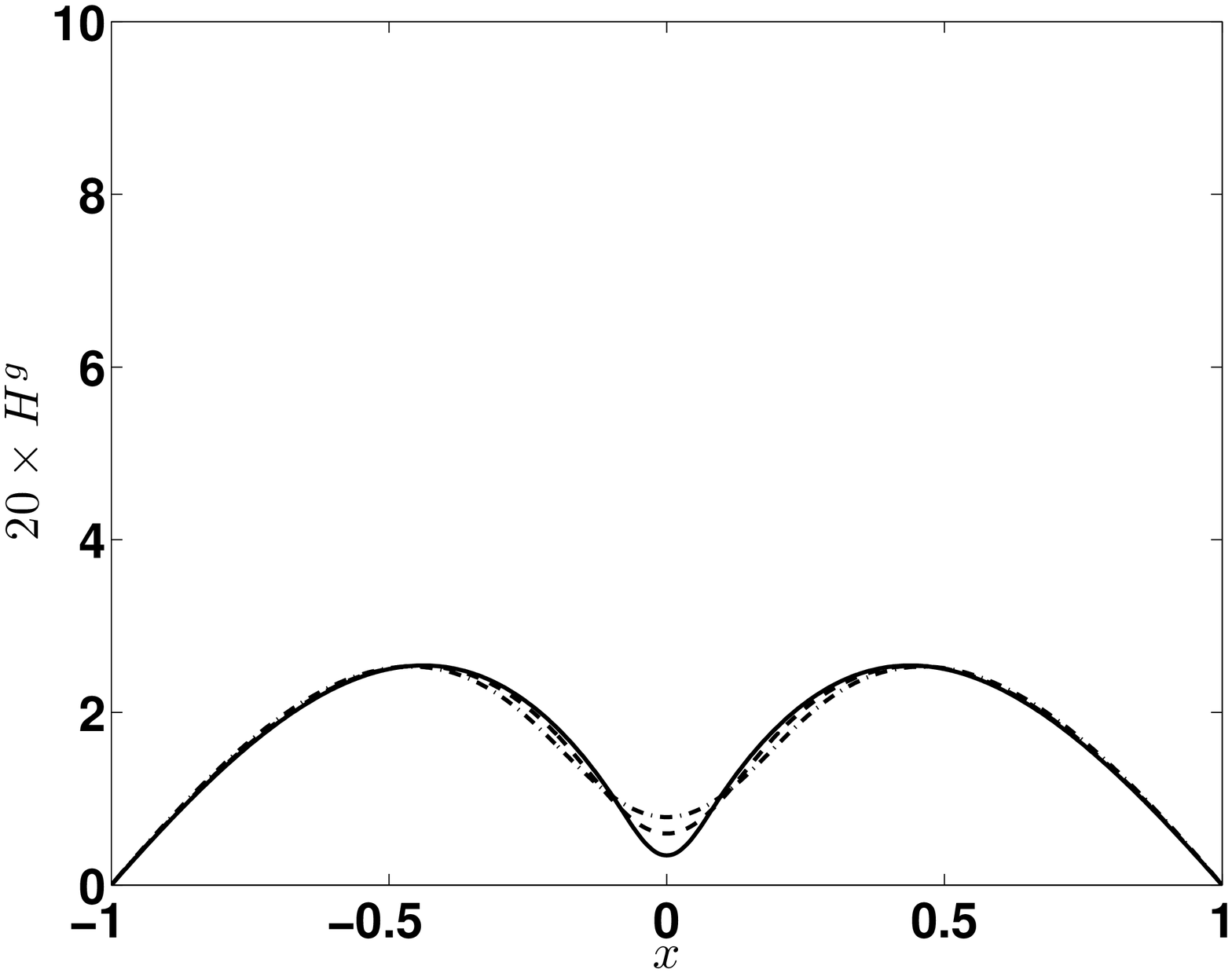}
\caption{\small (Color online) The Soft-Wall predictions for GPDs. Singlet-quark ({\bf upper panel}), Non-Singlet-quark ({\bf middle panel}) and gluon (amplified 20 times, {\bf lower panel}) GPDs at the low momentum scale $Q_0^2$ (see Eqs.(\ref{eq:sym}), (\ref{eq:antisym}) and (\ref{eq:symg})) at $t=-\Delta^2=0$, using the DDs  (\ref{eq:HqDD}). Full lines for $\xi = 0.1$, dashed  lines for $\xi = 0.2$ and dot-dashed lines for $\xi=0.3$. }
\label{fig:HS_HNS_Hg_Q02_SW}
\end{figure}

In the present Section the results for the chiral even distributions of the Soft-Wall model valid at $\xi=0$ as discussed in Section \ref{sec:Hu}, are generalized to $\xi > 0$ by means of the Double Distributions presented in the previous Section. They are defined in the different regions of the (generalized) $x$-values by the integrals (\ref{eq:DD2}) and the combinations (\ref{eq:sym}), (\ref{eq:antisym}) and (\ref{eq:symg}) (cfr. also ref.\cite{DD_Radyushkin2001}):
\bea 
&&H_{DD}^q(x,\xi,t=0)  \equiv  H_{DD}^q(x,\xi) = \nonumber \\ 
&& \theta(+\xi \leq x \leq +1)\,\int_{-{1-x \over 1+\xi}}^{{1-x \over 1-\xi}} d\alpha\,F^q(x-\xi \alpha,\alpha) + \nonumber \\
& + & \theta(-\xi \leq x \leq +\xi)\,\int_{-{1-x \over 1+\xi}}^{{1+x \over 1+\xi}} d\alpha\,F^q(x-\xi \alpha,\alpha) + \nonumber \\
& + & \theta(-1 \leq x \leq -\xi)\,\int_{-{1+x \over 1-\xi}}^{{1+x \over 1+\xi}} d\alpha\,F^q(x-\xi \alpha,\alpha) \nonumber \\
\label{eq:HqDD}
\eea
with
\bea 
&& H_{DD}^q(x,\xi,t=-\Delta^2) = H_{DD}^q(x,\xi,t=0)  \, x^{\Delta^2/(4 \alpha^2)}; \nonumber \\
&& x> 0. \nonumber
\eea
The results at low-momentum scale, $Q_0^2$, where the Soft-Wall model is supposed to be valid, are shown in Fig.\ref{fig:HS_HNS_Hg_Q02_SW} for three different values of the skewedness parameter $\xi = 0.1, 0.2$, and $0.3$ and invariant momentum $t=-\Delta^2=0$. The value of the low-momentum scale is identified by means of the momentum sum rule. In fact the number of particles are well defined at the initial scale (cfr. Eq.(\ref{eq:Nq})), and the momentum sum rule is not fulfilled by valence quarks only. As matter of fact one has:
\bea
&& \!\!\!\!\!\!\!\! \int dx\,x (H^u(x,\xi\!=\!0,t\!=\!0, Q_0^2) + H^d(x,\xi\!=\!0,t\!=\!0, Q_0^2))  =\nonumber \\
&=& \int x (u_V(x,Q_0^2)+d_V(x,Q_0^2) = (0.64+0.28) = 0.92. \nonumber \\
\label{eq:H_mom}
\eea
Differently from a quark model (relativistic or non-relativistic)
based on the presence of {\em only} valence quarks at the lowest scale, 
the holographic approach is intrinsically based on the QCD dynamics. The bound
system of valence quarks cannot share momentum among a pure three-quark system. 
The masses of the quarks are unknown and what is reproduced is
the spectrum of the system. The interpretation of the nucleon bound system implies the
presence of gluons exchanged among the valence quarks.
A natural consequence is an additional gluon distribution filling the gap to the total momentum.
A gluon distribution proportional to the valence densities at $Q_0^2$ (\`a la Gl\"uck, Reya, Vogt \cite{GluReyaVo}) can be a sensible choice 
\be
g(x,Q_0^2) = A_g \left[u_V(x,Q_0^2) + d_V(x,Q_0^2) \right], 
\label{eq:g_GR}
\ee
such that
\be
\int dx\, x (u_V + d_V + g) |_{Q_0^2} = 1,
\label{eq:mom_SR}
\ee
with $A_g = 0.091$ and $\int dx \, x g(x,Q_0^2) = 0.08$.
The (small) $H^g$ gluon distribution of Eq.(\ref{eq:symg}) as consequence of the density (\ref{eq:g_GR}), is shown in the lowest panel of Fig.\ref{fig:HS_HNS_Hg_Q02_SW}. The factor 20 is needed to make $H^g$ comparable with the results shown in the other panel of the same figure, $H^S$ and $H^{NS}$. 
The choice (\ref{eq:g_GR}) is only one of the possible choices one can make. 
One could assume a different parametrization of the gluon distribution (\ref{eq:g_GR}), 
and deduce a different behavior of the ({\em small}) $H^g$ component of Fig.\ref{fig:HS_HNS_Hg_Q02_SW}.
In the previous studies of GPDs within AdS/QCD no mention is made of the fact that the 
momentum sum rule is not satisfied, i.e. property (\ref{eq:H_mom}). The main reason to introduce here a conserving momentum sum rule like Eq.(\ref{eq:mom_SR}), is related to the possibility of a detailed investigation of the perturbative QCD evolution properties of the distributions. 
The simplified assumption made in Eq.(\ref{eq:g_GR}) is mostly connected to the fact that the perturbative evolution is dominated by the value of the moment carried by the gluon component rather than by the exact form of the distribution.  

In order to appreciate the role of the invariant momentum transfer $t = - \Delta^2$, the results of Fig.\ref{fig:HS_HNS_Hg_Q02_SW} valid at $t = 0$ are summarized in Fig.\ref{fig:HS_HNS_Hg_Q02_SW_t} and compared with the analogous predictions for $t = -0.5$ GeV$^2$. The SW model gives a non vanishing contribution to quark GPDs in the region $|x| < \xi$ at the lowest scale $Q_0^2$ without introducing discontinuities at $|x| = \xi$, the $\xi$-dependence is rather weak (cfr. Fig.\ref{fig:HS_HNS_Hg_Q02_SW}). One can check, in particular, that $H^S = H^{NS}$ at $x > \xi$, a peculiarity due to the absence of sea contribution at $Q_0^2$.}

\begin{figure}[tbp]
\centering\includegraphics[width=\columnwidth,clip=true,angle=0]{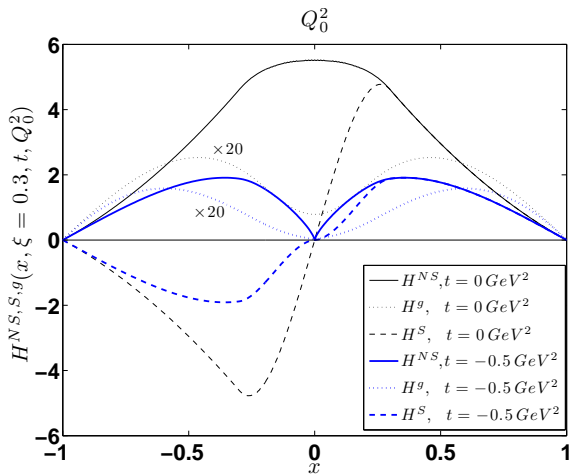}
\caption{\small (Color online) Comparison of the SW predictions for $H^{NS,S,g}$ at $\xi = 0.3$ and $t=0$ GeV$^2$, thin lines, results as in Fig.\ref{fig:HS_HNS_Hg_Q02_SW}, and $t = 0.5$ GeV$^2$, thick lines. The gluon distributions are amplified by a factor of 20. }
\label{fig:HS_HNS_Hg_Q02_SW_t}
\end{figure}

\section{\label{sec:conclusions}Conclusions and perspectives}

A study of GPDs within a general AdS/QCD framework, has been presented. Two main features have been emphasized and investigated in detail: 

\begin{itemize}

\item[i)] the role of the confining potential in the holographic coordinate as described within the Soft-Wall and within more general potential models; in particular the possibility of introducing high Fock
states in the calculation of GPDs. 

A method to study effects due to different confining potentials introduced to break conformal symmetry in the AdS/QCD approach to baryons has been proposed in Section \ref{sec:SWtoIR}. In several works devoted to the investigation of AdS/QCD wave functions for baryons, often the complementary aspect is stressed: the potentials must manifest iso-spectral properties and therefore their differences have to be adequately mitigated \cite{IR_isospectral}.  On the contrary the use of different (almost iso-spectral potentials) in calculating amplitude and responses in Deep Inelastic Scattering, can put in evidence relevant differences that can discriminate among them. The specific observables discussed in relation with Generalized Parton Distributions are a good example. The method implies the use of the Soft Wall solutions as a complete basis to solve more sophisticated potential models. 
The results are promising: the power of the holographic approach seems to be preserved and observables can be calculated following well established techniques. Higher Fock states can be accommodated showing their  relevance in the whole $x$-region.

\item[ii)] the extension of the GPDs results from AdS/QCD methods from the forward ($\xi=0$) to the off-forward region ($\xi>0$). The procedure used (Double - Distributions) enlarge the phenomenological domain of the GPDs predictions opening the concrete evaluation of the Single, Non-Singlet GPDs in the whole ($x$,$\xi$,t) domain. The procedure used identifies also the resolution scale of the results. The example developed is restricted to the Soft-Wall, but it is easily generalized to more complex confining potentials.

\end{itemize}

The calculated helicity independent and dependent GPDs show differences and properties to be further investigated in order to compare their predictions with the new generation of experimental data. In particular the important contributions due additional degrees of freedom like non-perturbative gluon and sea components should be further investigated together with a detailed analysis of the perturbative effects due to QCD evolution.  The elegance and the effectiveness of the AdS/QCD approaches has to be integrated in a complete predictive scheme for a large variety of observables in the perspectives of modeling the nucleon structure \cite{Modeling_nucleon}, work in that direction is in progress.

\begin{table}
\caption{The numerical values of the coefficients $a_{\nu n}^\pm$ for the variational expansion (\ref{eq:phi_exp}) in the case of maximum h.o. quanta $n_{max}=16$ and $\nu=4$ ($l_+ = \nu = 4$, $l_-=\nu+1=5$ and $\nu=5$ ($l_+ = \nu = 5$, $l_-=\nu+1=6$ ). The h.o. constants are fixed by the minimization procedure at $\alpha^+ = 2.35$ fm$^{-1}$ and $\alpha^- = 2.65$ fm$^{-1}$.}
\begin{center}
\begin{tabular}{|r|r|r|r|r|}
\hline
     &              &               &               &             \\ 
     $n$ & $a_{(\nu=4) n}^+$ & $a_{(\nu=4) n}^-$   & $a_{(\nu=5) n}^+$ & $a_{(\nu=5) n}^-$ \\     
 \hline
     &              &               &              &              \\
0  &  0.9592  &  0.8321  & 0.9497  & 0.8036  \\
1  &  -0.2753 & -0.4960 & -0.3038 & -0.5231 \\
2  &   0.0632 &  0.2278  & 0.0744  & 0.2577  \\
3  &  -0.0125 & -0.0914  &-0.0159 & -0.1100 \\
4  &  0.0023  &  0.0337  &  0.0031 & 0.0428  \\
5  & -4.0e-04& -0.0117  & -5.7e-04 & -0.0157\\
6  & 6.8e-05 & 0.0039   & 1.0e-04  & 0.0055  \\
7  &-1.1e-05 & -0.0013 & -1.7e-05 & -0.0018 \\
8  &  1.8e-06 & 3.9e-04 & 2.9e-06  & 6.0e-04 \\
9  & -2.8e-07 & -1.2e-04 & -4.7e-07&-1.9e-04\\
10&  4.3e-08 & 3.7e-05 & 7.6e-08  & 5.9e-05\\
11& -6.6e-09 & -1.1e-05 & -1.2e-08&-1.8e-05\\
12 &1.0e-09 & 3.2e-06  &1.9e-09 & 5.5e-06 \\
13 &-1.5e-10&-9.2e-07  &-2.9e-10& -1.6e-06 \\
14 & 2.9e-11 & 2.6e-07 & 4.1e-11 & 4.8e-07  \\
15 & 1.6e-11 & -7.2e-08& 1.3e-11 &-1.5e-07 \\
16 & -3.3e-11& 1.8e-08 & -5.0e-11& 4.7e-08\\
\hline
\end{tabular}
\end{center}
\label{tab:anl45}
\end{table}%

\appendix    					 	


\section{\label{sec:app3}Higher Fock states}

In this appendix some details of the method proposed in Section \ref{sec:SWtoIR} are illustrated.

The procedures can be generalized in order to accommodate higher Fock states. As discussed in Section \ref{sec:higherFock} the values of $\alpha^\pm$ remain the same also for the wave functions with $\nu=4$ and $\nu=5$ (as a numerical check has confirmed). The minimization produces the values of the coefficients shown in table \ref{tab:anl45}.

\begin{acknowledgements}
I would like to thank my colleagues J.-P. Blaizot and J.-Y. Ollitrault of the IPhT CEA-Saclay for the active  scientific atmosphere that has stimulated also the present study and for their continuous help.  A key mail exchange with  Zhen Fang is also gratefully acknowledged. I thanks S. Scopetta, V. Vento and M. Rinaldi for a critical reading of the manuscript and their fruitful collaboration. 
The last version of the present manuscript has been written in Perugia during a visiting period and I thanks the Sezione INFN and the Department of Physics and Geology for the warm hospitality and support. 
\end{acknowledgements}


\begin{thebibliography}{50}


\bibitem{DiehlPR2003}{\em Generalized Parton Distributions}
M. Diehl, \\
Phys. Rep. 388 (2003) 41-277.\\
hep-ph/0307382

\bibitem{generalGPDs}{\em Hard exclusive reactions and the structure of hadrons},\\
K. Goeke, M. V. Polyakov, and M. Vanderhaeghen, \\
Prog. Part. Nucl. Phys. {\bf 47}, 401 (2001); \\
arXiv:hep-ph/0106012\\
{\em Generalized Parton Distributions},\\
Xiangdong Ji,\\
Annu. Rev. Nucl. Part. Sci. {\bf 54}, 413 (2004); \\
{\em Unraveling hadron structure with generalized \\
parton distributions},\\
A. V. Belitsky and A. V. Radyushkin, \\
Phys. Rep. {\bf 418}, 1-387 (2005).\\
arXiv:hep-ph/0504030

\bibitem{GPDs_perp} {\em Impact Parameter Dependent Parton Distributions \\ 
and Off-Forward Parton Distributions for $\xi \to 0$},\\
M. Burkardt, \\
Phys. Rev. D {\bf 62}, 071503 (2000); {\bf 66}, 119903 (E) (2002); \\
arXiv:hep-ph/0005108\\
{\em Impact Parameter Space Interpretation for \\
Generalized Parton Distributions},\\
Int. J. Mod. Phys. A {\bf 18}, 173-208 (2003);\\
arXiv:hep-ph/0207047\\
{\em Gauge-Invariant Decomposition of Nucleon Spin \\ and its Spin-Off},\\
X. Ji,\\ 
Phys. Rev. Lett. {\bf 78}, 610-613 (1997); \\
arXiv:hep-ph/9603249\\
{\em Deeply Virtual Compton Scattering},\\
Phys. Rev. D {\bf 55}, 7114-7125 (1997).\\
arXiv:hep-ph/9609381

\bibitem{GPDs_FF} {\em Nucleon Form Factors from Generalized \\
Parton Distributions},\\
M. Guidal, M. V. Polyakov, A. V. Radyushkin, and M. Vanderhaeghen,\\
Phys. Rev. D {\bf 72}, 054013 (2005).\\
arXiv:hep-ph/0410251

\bibitem{GPDs_structure}{\em Generalized parton distributions and the structure \\
of the nucleon},\\
S. Boffi, B. Pasquini,\\
Riv. Nuovo  Cim. {\bf 30}, 387 (2007).\\
arXiv:hep-ph/07112625

\bibitem{GPDs_q}{\em Generalized parton distributions in \\ 
constituent quark models},\\
S. Scopetta, V. Vento,\\
Eur. Phys. J. A {\bf 16} 527-535 (2003).\\
arXiv:hep-ph/0201265\\
{\em Generalized parton distributions and composite constituent quarks},\\
Phys. Rev. D {\bf 69},  094004 (2004).\\
arXiv:hep-ph/0307150\\
{\em Helicity-dependent generalized parton distributions and \\ 
composite constituent quarks}\\
Phys. Rev. D {\bf 71},  014014 (2005).\\
arXiv:hep-ph/0410191

\bibitem{XJi_prize}{\em Proton Tomography Through Deeply Virtual \\
Compton Scattering},\\
Xiangdong Ji,\\
Feshbach Prize in theoretical nuclear physics, \\
talk at APS April Meeting, \\
Salt Lake City, April 18, 2016\\
arXiv:hep-ph/160501114

\bibitem{GPDs_CLAS_VCSc2015}{\em Single and double spin asymmetries for deeply \\
virtual Compton scattering measured with CLAS \\
and a longitudinally polarized proton target},\\
S. Pisano, A. Biselli, S. Niccolai, E. Seder, M. Guidal, M. Mirazita, the CLAS Collaboration,\\
Phys. Rev. D {\bf 91}, 052014 (2015).\\
arXiv:hep-ex/150107052

\bibitem{GPDs_CLAS_VMProd}{\em Exclusive $\pi^0$ electroproduction at $W>2$ GeV with CLAS},\\
I. Bedlinskiy et al. (CLAS Collaboration), \\
Phys. Rev. C {\bf 90}, 025205 (2014).\\
arXiv:hep-ex/14050988\\
{\em Measurement of Exclusive $\pi^0$ Electroproduction Structure Functions and their Relationship to Transversity GPDs},\\
Phys. Rev. Lett. {\bf 109}, 112001 (2012).\\
arXiv:hep-ex/12066355

\bibitem{GPDs_H1} {\em Measurement of Deeply Virtual Compton Scattering \\
at HERA},\\
C. Adloff et al. (H1 Collaboration), \\
Phys. Lett. B {\bf 517}, 47 (2001).\\
arXiv:hep-ex/0107005

\bibitem{GPDs_ZEUS}{\em Measurement of deeply virtual Compton scattering \\
at HERA},\\
S. Chekanov et al. (ZEUS Collaboration), \\
Phys. Lett. {\bf B 573}, 46 (2003).

\bibitem{GPDs_HERMES}{\em Measurement of the Beam-Spin Azimuthal Asymmetry \\
Associated with Deeply-Virtual Compton Scattering},\\
A. Airapetian et al. (HERMES Collaboration), \\
Phys. Rev. Lett. {\bf 87}, 182001 (2001).\\
arXiv:hep-ex/0106068

\bibitem{GPDs_HallA} {\em Deeply Virtual Compton Scattering off the neutron},\\
M. Mazouz et al. (Jefferson Lab Hall A Collaboration), \\
Phys. Rev. Lett. {\bf 99}, 242501 (2007).\\
nucl-ex/07090450

\bibitem{GPDs_COMPASS}{\em Feasibility study of deeply virtual Compton \\
scattering using COMPASS at CERN},\\
N. DÕHose, E. Burtin, P.A.M. Guichon, and J. Marroncle (COMPASS Collaboration), \\
Eur. Phys. J. A {\bf 19}, 47 (2004).\\
arXiv:hep-ex/0212047

\bibitem{GPDs_Lat} {\em Nucleon electromagnetic form factors from \\ 
lattice QCD using a nearly physical pion mass},\\
J. R. Green, J. W. Negele, A. V. Pochinsky, S. N. Syritsyn, M. Engelhardt, S. Krieg\\
Phys. Rev. D {\bf 90}, 074507 (2014).\\
arXiv:hep-lat/14044029\\
{\em $\langle x\rangle_{u-d}$ from lattice QCD at nearly physical quark masses},\\
G. S. Bali, S. Collins, M. Deka, B. Gl\"assle, M. G\"ockeler, J. Najjar,
A. Nobile, D. Pleiter, A. Sch\"afer, and A. Sternbeck, \\
Phys. Rev. D {\bf 86}, 054504 (2012);\\ 
arXiv:hep-lat/12071110 

\bibitem{AdS_Maldacena}{\em The Large N Limit of Superconformal Field \\
Theories and Supergravity},\\
J. M. Maldacena, \\
Adv. Theor. Math. Phys. {\bf 2}, 231 (1998).\\
arXiv:hep-th/9711200

\bibitem{AdS_Polyakov}{\em Gauge Theory Correlators from \\
Non-Critical String Theory},\\
S. S. Gubser, I. R. Klebanov and A. M. Polyakov, \\
Phys. Lett. {\bf B 428}, 105 (1998).\\
arXiv:hep-th/9802109

\bibitem{AdS_Witten1and2}{\em Anti De Sitter Space And Holography},\\ 
E. Witten, \\
Adv. Theor. Math. Phys. {\bf 2}, 253 (1998).\\
arXiv:hep-th/9802150\\
{\em Anti-de Sitter Space, Thermal Phase Transition, \\
And Confinement in Gauge Theories},\\
Adv. Theor. Math. Phys. {\bf 2}, 505 (1998).\\
arXiv:hep-th/9803131

\bibitem{top_down}{\em Exploring improved holographic theories for QCD: \\
Part I},\\
U. Gursoy and E. Kiritsis, \\
JHEP 0802, 032 (2008);\\
arXiv:hep-th/07071324\\
{\em Exploring improved holographic theories for QCD: \\
Part II},\\
U. Gursoy, E. Kiritsis and F. Nitti, \\
JHEP 0802, 019 (2008).\\
arXiv:hep-th/07071349 

\bibitem{hep-ph/1407.8131} {\em Light-Front Holographic QCD and Emerging \\ 
Confinement},\\
S. J. Brodsky, G. F. de Teramond, H. G. Dosch, J. Erlich,\\
Phys. Rept. {\bf 584}, 1 (2015). \\
arXiv:hep-ph/14078131

\bibitem{BT1}{\em AdS/CFT and light-front QCD, in Search
for the Totally Unexpected in the LHC Era},\\
S. J. Brodsky and G. F. de Teramond, \\
Proceedings of the International School of Subnuclear Physics, 
Vol. 45 (World Scientific Publishing Co., 2009).\\
arXiv:08020514 [hep-ph]

\bibitem{HW1} {\em QCD and a Holographic Model of Hadrons},\\
J. Erlich, E. Katz, D.T. Son and M.A. Stephanov,\\
Phys. Rev. Lett. {\bf 95}, 261602  (2005).\\
arXiv:hep-ph/0501128

\bibitem{HW2} {\em Chiral symmetry breaking from five dimensional spaces},\\
L. Da Rold and A. Pomarol,\\
Nucl. Phys. {\bf B721}, 79 (2005).\\
arXiv:hep-ph/0501218

\bibitem{SW1}{\em Linear Confinement and AdS/QCD},\\
A. Karch, E. Katz, D. T. Son and M. A. Stephanov,\\
Phys.Rev. D {\bf  74},  015005 (2006).\\
arXiv:hep-ph/0602229

\bibitem{ChiralTransition_SW2016}{\em Chiral Phase Transition in the Soft-Wall \\
Model of AdS/QCD},\\
K. Chelabi, Z. Fang, M. Huang, D. Li and Y.-L. Wu,\\
JHEP 04, 036 (2016). \\
arXiv:hep-ph/151206493

\bibitem{RealizationChiSB2016}{\em Realization of chiral symmetry breaking and restoration \\
in holographic QCD},\\
K. Chelabi, Z. Fang, M. Huang, D. Li and Y.-L. Wu,\\
Phys. Rev. D {\bf 93},  101901(R) (2016).\\
arXiv:hep-ph/151102721

\bibitem{IR_isospectral} {\em Family of dilatons and metrics for AdS/QCD models},\\
A. Vega and P. Cabrera,\\
Phys. Rev. D {\bf 93},  114026 (2016).\\
arXiv:hep-ph/160105999

\bibitem{IR_improvedN}{\em IR-improved Soft-Wall AdS/QCD \\
Model for Baryons},\\
Z, Fang, D. Li and Y. L. Wu, \\
Phys. Lett. {\bf B 754}, 343 (2016), \\
arXiv:hep-ph/160200379

\bibitem{IR_improvedM}{\em Infrared-Improved Soft-wall AdS/QCD \\ 
Model for Mesons},\\
Ling-Xiao Cui, Zhen Fang, Yue-Liang Wu,\\
Eur. Phys. J.  C  (2016) 76:22.\\
arXiv:hep-ph/13106487

\bibitem {IR_glueball}{\em Glueball spectra and Regge trajectories from \\
a modified holographic soft-wall model},\\
E. F. Capossoli and H. Boschi-Filho, \\
Phys. Lett. {\bf B 753}, 419 (2016). \\
arXiv:hep-ph/151003372

\bibitem{IR_dynamics}{\em Solution of the 5D Einstein equations in a dilaton \\
background model},\\
W. de Paula, T. Frederico, H. Forkel, M. Beyer,\\
Talk in Light Cone 2008 Mulhouse, \\
France, PoS LC2008:046, 2008.\\
arXiv:hep-ph/08102710\\
{\em Dynamical holographic QCD with area-law \\
confinement and linear Regge trajectories},\\
W. de Paula, T. Frederico, H. Forkel, M. Beyer,\\
Phys. Rev. D {\bf 79}, 075019 (2009);\\
arXiv:hep-ph/08063830\\
{\em Soft Walls in Dynamic AdS/QCD and \\
the Techni-dilaton},\\
N. Evans, P. Jones and M. Scott, \\
Phys. Rev. D {\bf 92}, 106003 (2015).\\
arXiv:hep-ph/150806540

\bibitem{hep-ph/11080346} {\em Dilaton in a soft-wall holographic approach to \\
mesons and baryons}\\
T. Gutsche, V. E. Lyubovitskij, I. Schmidt, A. Vega, \\
Phys. Rev. D {\bf  85}, 076003 (2012).\\
arXiv:hep-ph/11080346

\bibitem{AdS_DIS} {\em Hard Scattering and Gauge/String Duality},\\
J. Polchinski, M. J. Strassler,\\
Phys. Rev. Lett. {\bf  88},  031601 (2002)\\
arXiv:hep-th/0109174\\
{\em Deep Inelastic Scattering and Gauge/String Duality},\\
JHEP  0305, 012  (2003).\\
arXiv:hep-th/0209211

\bibitem{GPDs_HW}{\em Generalized parton distributions in an AdS/QCD \\
hard-wall model},\\
A. Vega, I. Schmidt, T. Gutsche and V. E. Lyubovitskij, \\
Phys. Rev. D {\bf 85}, 096004 (2012);
arXiv:hep-ph/12024806\\
{\em Investigating generalized parton distribution in \\
gravity dual},\\
R. Nishio, T. Watari, \\
Phys. Lett. {\bf B 707}, 362 (2012).\\
arXiv:hep-ph/11052907

\bibitem{GPDs_SW} {Generalized parton distributions in AdS/QCD},\\
A. Vega, I. Schmidt, T. Gutsche and V. E. Lyubovitskij, \\
Phys. Rev. D {\bf 83}, 036001 (2011);\\
arXiv:hep-ph/10102815\\
{\em Generalized Parton Distributions for the Proton  \\
in AdS/QCD},\\
Phys. Rev. D {\bf 88}, 073006 (2013); \\
arXiv:hep-ph/13075128\\
{\em Generalized Parton Distributions in \\ 
the Soft-Wall Model of AdS/QCD},\\
Neetika Sharma, \\
Phys. Rev. {\bf D 90}, 095024 (2014).\\
arXiv:hep-ph/14117486

\bibitem{GPDs_SW2}{\em Generalized parton distributions \\
for the proton in AdS/QCD},\\
D. Chakrabarti and C. Mondal,\\
Phys. Rev. D {\bf 88}, 073006 (2013).\\
arXiv:hep-ph/13075128

\bibitem{two_component1999}{\em Double distributions and evolution equations},\\
A.V. Radyushkin,\\
Phys. Rev. D {\bf 59}, 014030 (1998).\\
arXiv:hep-ph/9805342

\bibitem{hFock}{\em Nucleon structure including high Fock \\
states in AdS/QCD},\\ 
T. Gutsche, V. E. Lyubovitskij, I. Schmidt, A. Vega, \\
Phys. Rev. D {\bf 86},  036007 (2012),\\
arXiv:hep-ph/12046612

\bibitem{VQz}{\em Structure of vector mesons in a holographic \\
model with linear confinement},\\
H. R. Grigoryan and A. V. Radyushkin, \\
Phys. Rev. D {\bf 76}, 095007 (2007).\\
arXiv:hep-ph/07061543

\bibitem{BoffiPasquiniTraini1} {\em Linking generalized parton distributions \\ 
to constituent quark models},\\
S. Boffi, B. Pasquini and M. Traini,\\
Nucl. Phys. {\bf B649}, 243 (2003). \\
arXiv:hep-ph/0207340

\bibitem{FaccioliTrainiVento} {\em Polarized Parton Distributions and \\ 
Light-Front Dynamics},\\
P. Faccioli, M. Traini and V. Vento,\\
Nucl. Phys. {\bf A656}, 400 (1999).\\
arXiv:hep-ph/9808201 

\bibitem{GPDs_NonP-P}{\em Nonperturbative versus perturbative effects \\
in generalized parton distributions},\\
B. Pasquini, M. Traini and S. Boffi,\\
Phys. Rev. D {\bf 71},  034022 (2005).\\
arXiv:hep-ph/0407228

\bibitem{GPDs_cloud}{\em Virtual meson cloud of the nucleon \\
and generalized parton distributions},\\
B. Pasquini, S. Boffi,\\
Phys. Rev. D {\bf 73}, 094001 (2006).\\
arXiv:hep-ph/0601177 

\bibitem{GPDs_EV}{\em Next-to-leading order evolution of generalized \\
parton distributions for HERA and HERMES},\\
A. Freund and M. McDermott, \\
Phys. Rev. D {\bf 65}, 056012 (2002); \\
(E) {\bf 66}, 079903 (2002). \\
arXiv:hep-ph/0106115\\
{\em A next-to-leading order analysis of deeply virtual \\
Compton scattering},\\
Phys.Rev. D {\bf 65},  091901 (2002).\\
arXiv:hep-ph/0106124

\bibitem{GBE1}{\em Unified description of light- and strange-baryon spectra},\\
L.Ya. Glozman, W. Plessas, K. Varga and \\ 
R.F. Wagenbrunn, \\
Phys. Rev. D {\bf 58},  094030 (1998).\\
arXiv:hep-ph/9706507

\bibitem{Modeling_nucleon}{\em Modelling the nucleon structure},\\
M. Burkardt, B. Pasquini,\\
Eur. Phys. J. A (2016) 52:161.\\
arXiv:hep-ph/151002567

\bibitem{polynomiality1998}{\em Off-Forward Parton Distributions},\\
Xiangdong Ji,\\
J. Phys. G {\bf 24}, 1181 (1998).\\
arXiv:hep-ph/9807358

\bibitem{DD_Radyushkin2001}{\em Generalized Parton Distributions},\\
A.V. Radyushkin,\\
in {\em At the Frontier of Particle Physics/Handbook of QCD}, edited by M. Shifman (World Scientific, Singapore,2001).\\
arXiv:hep-ph/0101225

\bibitem{Traini_etal_PLB2017}{\em The effective cross section for double parton scattering within a holographic AdS/QCD approach},\\
M. Traini, M. Rinaldi, S. Scopetta and V. Vento,\\
Phys. Lett. {\bf B 768}, 270 (2017).\\ 
arXiv:hep-ph/1609.07242

\bibitem{GluReyaVo} M.Gl\"uck, E. Reya and A. Vogt, 
Z. Phys.  {\bf  C 67}, 433 (1995);\\
M. Gl\"uck, E. Reya and A. Vogt, Eur. Phys. J. {\bf C 5}, 461 (1998).



\end{thebibliography}
\end{document}